\documentclass[onecolumn,showpacs,preprintnumbers,amsmath,aps]{revtex4}
\usepackage{graphicx}
\usepackage{dcolumn}
\usepackage{bm}
\usepackage{color}
\usepackage{multirow}
\begin{document}
\def\eq#1{(\ref{#1})}
\def\fig#1{Fig.\hspace{1mm}\ref{#1}}
\def\tab#1{Tab.\hspace{1mm}\ref{#1}}
\title{Non-parametric application of Tsallis statistics\\ 
to systems consisting of $M$ hydrogen molecules}
\author{R. Szcz{\c{e}}{\'s}niak$^{\left(1\right)}$}
\email{szczesni@wip.pcz.pl}
\author{E. A. Drzazga$^{\left(1\right)}$}
\email{edrzazga@wip.pcz.pl}
\author{I. A. Domagalska$^{\left(1\right)}$}
\author{A. P. Durajski$^{\left(1\right)}$}
\author{M. Kostrzewa$^{\left(2\right)}$}
\affiliation{$^1$ Institute of Physics, Cz{\c{e}}stochowa University of Technology, Ave. Armii Krajowej 19, 42-200 Cz{\c{e}}stochowa, Poland}
\affiliation{$^2$ Institute of Physics, Jan D{\l}ugosz University in Cz{\c{e}}stochowa, Ave. Armii Krajowej 13/15, 42-200 Cz{\c{e}}stochowa, Poland}
\date{\today} 
\begin{abstract}
We have determined the entropy, the total energy, and the specific heat of the systems consisting of $M\geq 3$ hydrogen molecules. 
The calculations were conducted in the framework of the nonextensive Tsallis statistics. The relation between $M$ and the entropic index $q$ is given by $q = 1 + 1/M$, which results from the fact that the temperature of the nanosystems fluctuates around the temperature of the reservoir (Wilk and W{\l}odarczyk, Phys. Rev. Lett. {\bf 84}, 2770 (2000)). The electron energy states of the hydrogen molecule have been determined with the help of the Hubbard Hamiltonian, which models all two-body interactions. The Hubbard Hamiltonian integrals have been calculated by using the variational method, whereas the Wannier function has been associated with $1s$ Slater-type orbitals. We have included the contributions to the energy of the hydrogen molecule coming from the oscillatory (either harmonic or anharmonic), rotational and translational degrees of freedom. In addition, we have investigated the impact of the external force ($F$) or the magnetic field ($h$) on the thermodynamic parameters of the systems. In each case the noticeable deviation from the results of the classical statistical physics can be observed for the systems consisting of $M<M_{c}\sim 10^{3}$ molecules.  
\end{abstract}
\pacs{05, 05.30.Ch, 31.15.-p, 31.15.A-, 31.15.aq}
\maketitle
\noindent{\bf Keywords:} Nonextensive Tsallis statistics,Small number of elements, Hydrogen molecule, {\it Ab initio} calculations\\ 
%

\section{Introduction}

The nonextensive (pseudo-additive) statistics was initially proposed by Tsallis \cite{Tsallis1988A, Tsallis1998A, Curado1991A, Curado1992A}.
This is because the standard method based on the Boltzmann-Gibbs-Shannon (BGS) statistics cannot properly deal with 
the systems where the $M$-body entropy is not proportional to $M$. For example, the nonextensivity has been realized in the $d$-dimensional systems with the long-range interactions characterized by the potential energy $\sim 1/r^{\gamma}$ ($\gamma>0$). The typical physical examples are: the nonionized hydrogen atom gas in the free space and the gravitation systems ($\gamma=1$ and $d=3$) \cite{Lucena1995A, Padmanabhan1990A}.
Let us notice that for $\gamma\leq d$ the standard BGS partition function diverges at the long distance limit. The nonextensivity of the entropy is also considered in the multi-fractal systems \cite{Tsallis2004A}, and in the small-scale systems with the fluctuation of the temperature or the energy dissipation \cite{Schmidt2001A}. The Tsallis entropy is defined in the form \cite{Tsallis2004A, Tsallis2010A, Ferri2005A}:
\begin{equation}
\label{r01a}
S_{q}=\frac{k_{B}}{q-1}\left(1-\sum_{j}p^{q}_{j}\right)=-k_{B}\sum_{j}p^{q}_{j}\ln_{q}\left(p_{j}\right),
\end{equation}
where $q$ is indicating the entropic index, $k_{B}$ represents the Boltzmann constant, $p_{j}$ are the probabilities of the microscopic configurations:
\begin{equation}
\label{r02a}
p_{j}=\frac{\exp_{q}\left(-\beta' E_{j}\right)}{\sum_{n}\exp_{q}\left(-\beta' E_{n}\right)}.
\end{equation}
The $q$-logarithm is given by $\ln_{q}\left(x\right)=\left(x^{1-q}-1\right)/\left(1-q\right)$, while 
$\exp_{q}\left(x\right)=\left[1+\left(1-q\right)x\right]^{\frac{1}{1-q}}$ denotes the $q$-exponent, which is the function inverse to the $q$-logarithm. 
Note that the $q$-exponent is the solution of the differential equation $dy/dx=y^{q}$, whereas $y\left(0\right)=1$. The quantity $\beta'=1/k_{B}T'$ represents the intermediate temperature, which is used for the computational sake:
\begin{equation}
\label{r03a}
\beta'=\frac{\beta}{\sum_{j}p^{q}_{j}+\left(1-q\right)\beta U_{q}},
\end{equation}
and $\beta=1/k_{B}T$. The Tsallis entropy is so general that if we accept $q\rightarrow 1$, it reduces to the standard BGS statistics 
($S=-k_{B}\sum_{j}p_{j}\ln p_{j}$). The nonextensive feature of $S_{q}$ becomes evident when we apply the law of the entropy submission 
for two statistically independent systems $A$ and $B$ ($p_{ij}\left(A+B\right)=p_{i}\left(A\right)p_{j}\left(B\right)$):
\begin{equation}
\label{r04a}
S_{q}\left(A+B\right)=S_{q}\left(A\right)+S_{q}\left(B\right)+\left(1-q\right)S_{q}\left(A\right)S_{q}\left(B\right).
\end{equation}
Depending on the value of $q$ we can talk about subextensivity ($q>1$) or superextensivity ($q<1$). 
For the other properties of $q$-entropy we have: $S_{q}$ is non-negative, extremal for equiprobability, and is concave (convex) if $q>0$ ($q<0$). The Tsallis entropy satisfies the $H$ theorem \cite{Mariz1992A, Ramshaw1993A, Ramshaw1993B}, {\it i.e.}, $dS_{q}/dt\geq 0$ ($\leq 0$) if $q>0$ ($q<0$). The $q$-statistics leave invariant form for all values of $q$ (the Legendre transform structure of thermodynamics \cite{Curado1991A, Curado1992A}, the Ehrenfest theorem and the von Neumann equation \cite{Plastino1993A, Plastino1994A}, as well as the Onsager reciprocity theorem). 
They satisfy Jaynes information theory duality relations \cite{Plastino1993A, Plastino1994A}, generalize the Langevin and the Fokker-Planck equations \cite{Stariolo1994A}, the quantum statistics \cite{Buyukkili1993A, Buyukkili1995A}, and the fluctuation-dissipation theorem \cite{Chame1994A}. 
The presented formalism assumes two constraints on the system, the normalization $\sum_{j}p_{j}=1$, and the conservation of the total energy:
\begin{equation}
\label{r05a}
U_{q}=\frac{\sum_{j}p^{q}_{j}E_{j}}{\sum_{j}p^{q}_{j}}.
\end{equation}

The main aim of this paper is to determine the limit of applicability of the conventional Boltzmann-Gibbs-Shannon statistics 
($q=1$) in the case of systems consisting of small number of the hydrogen molecules. To this end, in the framework of the non-parametric theory we have calculated the electron energy states of the single hydrogen molecule. We took into account all two-body electron interactions, wherein we subordinated the Hamiltonian coefficients on the distance between protons. In the present paper, we have  considered the oscillatory (either harmonic or anharmonic), rotational and translational states of the hydrogen molecule. We take also into account the constant force acting on the molecule or the constant external magnetic field. 

In addition, we adopted the natural assumption that the temperature $T$ of the small-scale systems fluctuates around the temperature $T'$ of the reservoir because of the smallness of the system. More mathematically, let $E$ stand for the internal energy for the small system, which is immersed in the large reservoir with the energy $E'$. Then the BGS distribution should be averaged over the fluctuating inverse temperature $\beta$ \cite{Wilk2000A, Beck2002A, Rajagopal2004A, Hasegawa2006A}:
\begin{equation}
\label{r06a}
\int_{0}^{+\infty}d\beta e^{-\beta E}f\left(\beta,\beta'\right)=\left[1-\left(1-q\right)\beta'E\right]^{\frac{1}{1-q}},
\end{equation}
where $q = 1 + 1/M$, and:
\begin{equation}
\label{r07a}
f\left(\beta,\beta'\right)=\frac{1}{\Gamma\left(M\right)}\left(\frac{M}{\beta'}\right)^{M}\beta^{M-1}\exp\left(-\frac{\beta}{\beta'}\right).
\end{equation}
The formula \eq{r06a} is valid for arbitrary energy $E$ and thus of the great importance. It can be derived from the integral representation of the gamma-function \cite{Wilk2000A, Beck2001A}. Let us note that the function $f\left(\beta,\beta'\right)$ represents the probability density of the $\chi^{2}$ distribution. The right-hand side of Eq. \eq{r06a} is the generalized Boltzmann factor and it can be obtained by extremizing $S_{q}$. 
In other words, if we consider the small hydrogen molecule system, then the Tsallis distribution function is the consequence of the integrating over all possible fluctuating inverse temperatures $\beta$, provided $\beta$ is $\chi^{2}$ distributed \cite{Beck2002A}.
It is worthwhile to point that in the case of the thermodynamic limit (when the number of the items in the system goes to infinity) 
considerations related to the systems of the non-interacting molecules can be carried out with the aid of the Boltzmann distribution starting from the principle of the equal {\it a priori} probability without considering the statistical ensemble or assuming the particular form of the entropy \cite{McQuarrie1997A}. In the following part of the paper all designated thermodynamic quantities (the entropy, the total energy, and the specific heat) will be openly dependent on $M$ instead of $q$.

In our opinion, the presented results may have the fundamental significance not only for the system composed of the hydrogen molecules, but also for many other physical systems consisting of the relatively small number of the elements. The results obtained by us suggest that between the family of the systems described directly by the Newton or Schr{\"o}dinger equation (the systems composed of the several elements) and the family of the systems, for which parameters should be determined using the laws of the statistical physics, there is the clearly distinguished class of the systems, for which the thermodynamic parameters should be calculated using methods of the non-extensive statistical physics. Note that from the point of view of the current research trends, the particular attention should be paid to the electron or spin nanosystems, which usually contain from several to several hundred components \cite{Hill2001A, Hill2001B, Hill2001C, Rajagopal2004A, Szczesniak2012K, Biborski2016A}. Observed by us predictions' deviations basing on the Tsallis statistics, in relation to the results obtained with the aid of the BGS statistical mechanics, force the need to verify the currently used calculation methods for determining thermodynamic properties in this type of the physical systems, which is consistent with the conclusions in the publications \cite{Hartmann2004A, Hartmann2004B, Hartmann2004C, Hartmann2006A, Boer2008A}. Of course, it should be emphasized that in the presented paper we analyze the canonical system of the non-interacting hydrogen molecules. In many other cases, it is necessary to consider the interactions between the components of the tested system. As the result, the non-trivial modifications of the predictions obtained for the non-interacting system may occur, which is the new interesting field of the research. In this context, it is worth paying reader's particular attention to the low-temperature properties of the many physical systems. These include issues related to the thermodynamics of the electronic system in the crystals, where the usefulness of the Tsallis statistics is suggested for the description of the superconducting condensate \cite{Nunes2001A, Nunes2002A, Ribeiro2012A}. Equally interesting are the considerations for the spin systems \cite{Cavallo2002A, Nobre2003A, Saguia2009A}, the inhomogeneous magnetic systems \cite{Reis2006A, Reis2006B} or the quantum dots \cite{Khordad2013A}. Of course, you cannot forget also about the Bose-Einstein condensation analyzed in the context of the Tsallis statistics \cite{Miller2006A}. However, this is the case where interaction between elements of the system is not taken into account (whereas the symmetry of the wave function plays the fundamental role).

The numerical calculations related to the Tsallis statistics were made on the basis of the algorithm according to Lim and Penn \cite{Lima1999A}. 
The mentioned algorithm has the following form: 
(1) compute the quantities $y_{j}=\left[1-\left(1-q\right)\beta'E_{j}\right]$ for all of $j$,
(2) if $y_{j}<0$ then $y_{j}=0$ (the cut-off condition),
(3) compute the partition function $Z_{q}=\sum_{j}y_{j}^{\frac{1}{1-q}}$,
(4) obtain $U_{q}(\beta')$ and other thermodynamic quantities,
(5) obtain $\beta(\beta')$ from Eq.~\eq{r03a}.

\section{The {\it ab initio} description of the hydrogen molecule}

The electron-proton energy of the hydrogen molecule ($E_{ep}$), under the external force ($F$) can be expressed by the formula:
\begin{eqnarray}
\label{r01b}
E_{ep}=E_{e}+E_{p}+E_{F},
\end{eqnarray}
where $E_{e}$ denotes the energy of the electron eigenstate, $E_{p}$ represents the energy of the proton-proton repulsion 
$E_{p}=2/R$, where $R$ is the distance between the protons. The last term is given by $E_{F}=FR$. From the formula \eq{r01b} results the physical meaning of the energy $E_{ep}$. It is equal to the sum of the internal energy ($E_{in}=E_{e}+E_{p}$), which is needed to create the hydrogen molecule, when the molecule is in the vacuum, and the product $FR$, which represents the work to be done to compress the molecule. Let us note that by explicitly defining dependence $E_{ep}\left(R\right)$, it is possible to calculate the contributions to the energy $E_{in}$ originating from the oscillating and the rotating motion of the proton system ($E_{o}$ and $E_{r}$). In addition, the molecule has three translational degrees of freedom with which energy $E_{t}$ is associated. The method of the determining the value of $E_{o}$, $E_{r}$ and $E_{t}$ has been described at the end of the current chapter. However, it is worth paying attention to the scale of the absolute values of the individual energies. The energy $E_{ep}$ in the equilibrium state we count in Rydberg, $E_{o}$ in the ground state is of the order $10^{-2}$~Ry, the difference in the energy $E_{r}$ between the ground state and the first excited one is the contribution of about $10^{-4}$~Ry. Even smaller values of the energy are obtained for the translational states. The applicability of the classic formulas on the energies $E_{p}$ and $E_{F}$ in the expression \eq{r01b} is based on the Born-Oppenheimer approximation \cite{Born1927A, Born1928A}, in the framework of which it is assumed that the full wave function of the system can be shown as the product of the functions of the electron subsystem ($\Phi$) and the protons ($\Psi$). Due to the fact that the mass of the proton is about $1800$ times greater than the mass of the electron, the proton position vectors are included in $\Phi$ parametrically. For the same reason, the wave equation on $\Psi$ is omitted, and the energy $E_{p}$ is calculated classically. Our calculations did not include the non-adiabatic terms that give the fixes of the order not less than $\sigma^{1/4}$, where $\sigma$ is the ratio of the electron to the proton mass \cite{Born1954A}.  

The energy of the electron states of the molecule has been determined with the help of the full Hubbard Hamiltonian ($\hat{H}$) in the second quantization notation. The method of the second quantization is equivalent to the description carried out with the use of Schr{\"o}dinger equation 
\cite{Schrodinger1926A, Schrodinger1926B, Schrodinger1926C, Schrodinger1926D}. It has been described in detail in the papers \cite{Spalek2015A, Kadzielawa2014A, Acquarone1998A}. In particular, the energy operator takes the following form:
\begin{eqnarray}
\label{r02b}
\hat{H}&=&\varepsilon\left(\hat{n}_1+\hat{n}_2\right)
       +t\sum_{\sigma}\left(\hat{c}_{1\sigma}^{\dag} \hat{c}_{2\sigma}+\hat{c}_{2\sigma}^{\dag}\hat{c}_{1\sigma}\right)\\ \nonumber
       &+&U\left(\hat{n}_{1\uparrow}\hat{n}_{1\downarrow}+\hat{n}_{2\uparrow}\hat{n}_{2\downarrow}\right)+
       K\hat{n}_1\hat{n}_2 \\ \nonumber 
       &-&J\left(2\hat{\mathbf{S}}_1\hat{\mathbf{S}}_2-\frac{1}{2}\hat{n}_1\hat{n}_2\right) 
        +J\left(\hat{c}_{1\uparrow}^{\dag}\hat{c}_{1\downarrow}^{\dag} \hat{c}_{2\downarrow}\hat{c}_{2\uparrow}+h.c.\right)\\ \nonumber 
       &+& V\sum_{\sigma}\left[ \left(\hat{n}_{1-\sigma}+\hat{n}_{2-\sigma} \right) \left(\hat{c}_{1\sigma}^{\dag} \hat{c}_{2\sigma}
       +\hat{c}_{2\sigma}^{\dag} \hat{c}_{1\sigma}\right) \right]\\ \nonumber
       &-&2h\left(\hat{S}^{z}_{1}+\hat{S}^{z}_{2}\right). 
\end{eqnarray}
The operator \eq{r02b} models all inter-electron interactions only in the diatomic systems. In molecules with more than two atoms, there will be additional types of the interactions not included in the formula \eq{r02b}. They should be determined directly from the general Hamiltonian for the electronic subsystem recorded with the help of the fermion field operators. 

The symbols appearing in the formula \eq{r02b} have the following meanings: $\hat{c}^{\dag}_{j\sigma}$ and $\hat{c}_{j\sigma}$ are the creation and annihilation operators of the electron state on $j$-th proton. 
The number operator is given by $\hat{n}_{j}=\sum_{\sigma}\hat{n}_{j\sigma}=\hat{c}_{j\sigma}^{\dag}\hat{c}_{j\sigma}$. 
$\hat{{\bf S}}_{j}$ represents the spin operator. Hence, 
$\hat{\mathbf{S}}_1\hat{\mathbf{S}}_2=\frac{1}{2}\left(\hat{S}^{+}_{1}\hat{S}^{-}_{2}+\hat{S}^{-}_{1}\hat{S}^{+}_{2}\right)+\hat{S}^{z}_{1}\hat{S}^{z}_{2}$, where $\hat{S}^{+}_{j}=\hat{c}^{\dag}_{j\uparrow}\hat{c}_{j\downarrow}$, $\hat{S}^{-}_{j}=\hat{c}^{\dag}_{j\downarrow}\hat{c}_{j\uparrow}$, and 
$\hat{S}^{z}_{j}=\frac{1}{2}\left(\hat{n}_{j\uparrow}-\hat{n}_{j\downarrow}\right)$. Other coefficients are defined by the following integrals:
\begin{eqnarray}
\label{r03b}
\varepsilon &=&
\int d^{3}{\bf r}\Phi_{1}\left({\bf r}\right)\left[-\nabla^{2}-\frac{2}{|{\bf r}-{\bf R}|}\right]\Phi_{1}\left({\bf r}\right),\\ \nonumber
t &=&
\int d^{3}{\bf r}\Phi_{1}\left({\bf r}\right)\left[-\nabla^{2}-\frac{2}{|{\bf r}-{\bf R}|}\right]\Phi_{2}\left({\bf r}\right),\\ \nonumber
U &=&
\int\int d^{3}{\bf r}_{1}d^{3}{\bf r}_{2}\Phi^{2}_{1}\left({\bf r}_{1}\right)\frac{2}{|{\bf r}_{1}-{\bf r}_{2}|}\Phi^{2}_{1}\left({\bf r}_{2}\right),\\ \nonumber
K &=&
\int\int d^{3}{\bf r}_{1}d^{3}{\bf r}_{2}\Phi^{2}_{1}\left({\bf r}_{1}\right)\frac{2}{|{\bf r}_{1}-{\bf r}_{2}|}\Phi^{2}_{2}\left({\bf r}_{2}\right),\\ \nonumber
J &=&
\int\int d^{3}{\bf r}_{1}d^{3}{\bf r}_{2}\Phi_{1}\left({\bf r}_{1}\right)\Phi_{2}\left({\bf r}_{1}\right)\frac{2}{|{\bf r}_{1}-{\bf r}_{2}|}\Phi_{1}\left({\bf r}_{2}\right)\Phi_{2}\left({\bf r}_{2}\right),\\ \nonumber
V &=&
\int\int d^{3}{\bf r}_{1}d^{3}{\bf r}_{2}\Phi^{2}_{1}\left({\bf r}_{1}\right)\frac{2}{|{\bf r}_{1}-{\bf r}_{2}|}\Phi_{1}\left({\bf r}_{1}\right)\Phi_{2}\left({\bf r}_{2}\right). \nonumber
\end{eqnarray}

The quantity $\varepsilon$ determines the energy of the single electron orbital, $t$ is the hopping integral, $U$ models the on-site Coulomb repulsion between the electrons, $K$ is the energy of the interstitial Coulomb repulsion. The symbol $J$ denotes the integral of the exchange, wherein the first term with $J$ in the formula \eq{r02b} presents the full expression on the impact of the exchange (the Dirac exchange operator). The second term with $J$ represents the hopping of the singlet pairs of the electrons. $V$ it is the energy of the correlated electron hopping. In the integrals collected in the formulas \eq{r03b} there are no spin components of the wave function ($\varphi_{s}$) due to the possibility of the choosing the global spin quantization axis ($\varphi^{\dag}_{s}\varphi_{s}=1$). The symbol $h$ in the formula \eq{r02b} is the applied magnetic field.

We have chosen the Wannier function as:
\begin{equation}
\label{r04b}
\Phi_{j}\left({\bf r}\right)= A\left[\phi_{j}\left({\bf r}\right)-B\phi_{l}\left({\bf r}\right)\right],
\end{equation}
wherein the parameters ensuring the normalization are of the form:
\begin{eqnarray}
\label{r05b}
A&=& \frac{1}{\sqrt{2}}\sqrt{\frac{1+\sqrt{1-S^{2}}}{1-S^{2}}},\\ 
B&=& \frac{S}{1+\sqrt{1-S^{2}}}.
\end{eqnarray}
The atomic overlap (S) should be calculated with the help of the formula $S=\int d^{3}{\bf r}\phi_{1}\left({\bf r}\right)\phi_{2}\left({\bf r}\right)$, where $1s$ Slater-type orbital can be written as $\phi_{j}\left({\bf r}\right)= \sqrt{\alpha^{3}/\pi}\exp\left[-\alpha|{\bf r}-{\bf R}_{j}|\right]$, 
$\alpha$ is the inverse size of the orbital.

\begin{figure} 
\includegraphics[width=0.5\columnwidth]{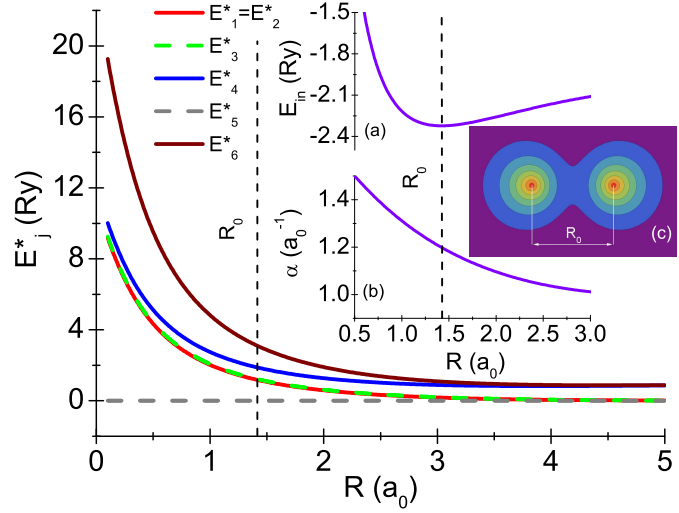}
\caption{The electron energy states of the hydrogen molecule ($E^{\star}_{j}=E_{j}-E_{5}$) as a function of the distance between the protons 
        ($F=0$, $h=0$). 
         The insertion presents: (a) the energy $E_{in}$ as a function of the distance, 
                                 (b) the inverse size of the orbital $\alpha$ as a function of the distance, and
                                 (c) the distribution of the electron charge density in the equilibrium condition.}
\label{f1}
\end{figure}

We have conducted the analytical diagonalization of the Hamiltonian \eq{r02b}. As the result we have obtained following formulas:
\begin{eqnarray}
\label{r06b}
E_1&=&-2h-J+K+2\varepsilon,
\end{eqnarray}
\begin{eqnarray}
\label{r07b}
E_2&=&2h-J+K+2\varepsilon,
\end{eqnarray}
\begin{eqnarray}
\label{r08b}
E_3&=&J+K+2\varepsilon,
\end{eqnarray}
\begin{eqnarray}
\label{r09b}
E_4&=&-J+U+2\varepsilon,
\end{eqnarray}
\begin{eqnarray}
\label{r10b}
E_5&=&\dfrac{1}{2}( K+U+4\varepsilon-\sqrt{N}),
\end{eqnarray}
\begin{eqnarray}
\label{r11b}
E_6&=&\dfrac{1}{2}\left( K+U+4\varepsilon+\sqrt{N}\right), 
\end{eqnarray}
where $N=\left(U-K\right)^{2}+16\left(t+V\right)^{2}+4J\left(U-K+J\right)$.

On the basis of the dependence $E_{ep}$ from $R$, it can be additionally calculated the vibrational energy levels. In the harmonic approximation, the potential is given by:
\begin{equation}
\label{r12b}
V_{\rm H}\left(R\right)=E_{0}+\frac{1}{2}k_{\rm H}\left(R-R_{0}\right)^{2},
\end{equation}
where: $E_{0}=E_{ep}\left(R_{0}\right)$, and $k_{\rm H}=\left[d^{2}E_{ep}\left(R\right)/dR^{2}\right]_{R=R_{0}}$ (the symbol $R_{0}$ represents the equilibrium distance). The quantum oscillator's energy possesses the well know form: 
\begin{equation}
\label{r13b}
E^{\rm H}_{o}=\omega^{\rm H}_{0}\left(n+1/2\right),
\end{equation}
wherein: $\omega^{\rm H}_{0}=\sqrt{k_{\rm H}/m'}$, and $n=0,1,2,...$ . The symbol $m'$ represents the reduced mass of the proton subsystem: 
$m'=m_{p}/2=918.076336$ ($m_{p}$ is the proton mass).

The harmonic description of the vibrational levels does not take into consideration the states close to the dissociation. 
More accurate calculations are based on the Morse potential ($V_{\rm Mo}$), which approximates the general form of the intermolecular energy curve. 
Note that the Morse curve represents the anharmonic potential also about the equilibrium proton separation. For $F=0$, the Morse potential can be written as:
\begin{equation}
\label{r14b}
V_{\rm Mo}\left(R\right)=E_{0}+E_{\rm D}\left[1-\exp\left(-\alpha_{\rm Mo}\left(R-R_{0}\right)\right)\right]^{2},
\end{equation}
where $E_{\rm D}$ is the dissociation energy of the molecule measured from the minimum value of $V_{\rm Mo}\left(R\right)$, and $\alpha_{\rm Mo}$
represents the measure of the curvature of the potential at its minimum. The force constants $k_{\rm Mo}$ is given by the formula
$k_{\rm Mo}=\left[d^{2}V_{\rm Mo}\left(R\right)/dR^{2}\right]_{R=R_{0}}$. Additionally, we have introduced the Morse energy 
$\omega^{\rm Mo}_{0}=\sqrt{k_{\rm Mo}/m'}$. The energy levels can be calculated based on the formula:
\begin{equation}
\label{r15b}
E^{\rm Mo}_{o}=\omega^{\rm Mo}_{0}\left(n+1/2\right)+\left((\omega^{\rm Mo}_{0})^{2}/4E_{D}\right)\left(n+1/2\right)^{2}.
\end{equation}
In the case of $F>0$, qualitatively the course of the curve $E_{ep}\left(R\right)$ is reproduced by the following formula:
\begin{equation}
\label{r16b}
V^{+}\left(R\right)=V_{\rm Mo}\left(R\right)+\frac{E_{\rm D}}{R}\left[1-\exp\left(-2.5\alpha_{\rm Mo}\left(R-R_{0}\right)\right)\right]^{2}.
\end{equation}

The rotational energy of the hydrogen molecule should be calculated using the formula:
\begin{equation}
\label{r17b}
E_{r}=B_{0}l\left(l+1\right),
\end{equation}
where $B_{0}$ is the rotational constant: $B_{0}=1/m' R^{2}_{0}$ and $l=0,1,2,...$ . In the presented description, we used the approximation which assumes that oscillations of the hydrogen molecules do not affect the rotation, and the rotation of the molecule has no effect on the oscillations. 

We estimated the translational energy of the hydrogen molecule basing on the formula: 
\begin{equation}
\label{r18b}
E_{t}=\frac{k^{2}}{16m_{p}},
\end{equation}
where ${\bf k}=(\pi k_{x}/L,\pi k_{y}/L,\pi k_{z}/L)$, $k_{j}=1,2,...$ .

\begin{figure} 
\includegraphics[width=0.5\columnwidth]{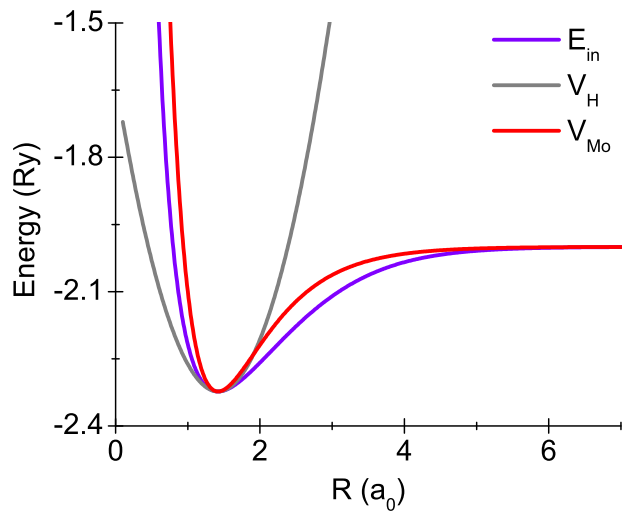}
\caption{The energy of the electron-proton system and the potentials $V_{\rm H}$ and $V_{\rm Mo}$ depending on the distance between the protons.}
\label{f2}
\end{figure}

We have determined the dependence of the energy of the electron states of the hydrogen molecule on the distance between the protons. 
We adopted the agreement that the level of zero coincides with the state of the lowest energy ($E_{min}=E_{5}$). The results are presented in \fig{f1}, whereas the vertical dashed line represents the equilibrium distance ($R_{0}=1.41968$~${\rm a_{0}}$), which corresponds to $\alpha_{0}=1.199205$~${\rm a_{0}^{-1}}$. The full course of the dependence of the total energy ($E_{in}$) on $R$ can be traced in the inset (a). The inset (b) characterize values of the inverse size of the orbital, and the inset (c) shows the distribution of the electron charge density ($\rho\left({\bf r}\right)=\sum_{j}\Phi^{\star}_{j}\left({\bf r}\right)\Phi_{j}\left({\bf r}\right)$) in the hydrogen molecule for 
$R=R_{0}$. Note that we have obtained results that very accurately reproduce the most advanced calculations presented in the branch literature. 
For example, the virtually exact solution given by Ko{\l}os and Wolniewicz has the form: 
$R_{0}=1.3984$~${\rm a_{0}}$ and $E_{0}=-2.349$~Ry \cite{Kolos1964A}, \cite{Kolos1968A} (in our case $E_{0}=-2.323011$~Ry). 
Very similar results were obtained also by K{\c{a}}dzielawa {\it et al.} \cite{Kadzielawa2014A}: $R_{0}=1.43042$~${\rm a_{0}}$ and $E_{0}=-2.29587$~Ry. 
Let us notice that the physical parameters of the hydrogen molecule can be also calculated with the help of the software available online, however results achieved in that way are less accurate. For example, the value of $E_{0}$ obtained in the Mopac software package \cite{Mopac} differs from the results of Ko{\l}os and Wolniewicz by $12$~\%. Additionally, in \fig{f2} we have drawn the course of the internal energy in the harmonic approximation, and the course reproduced by means of the Morse curve.

\section{THE NONEXTENSIVE THERMODYNAMICS}

\begin{figure} 
\includegraphics[width=0.49\columnwidth]{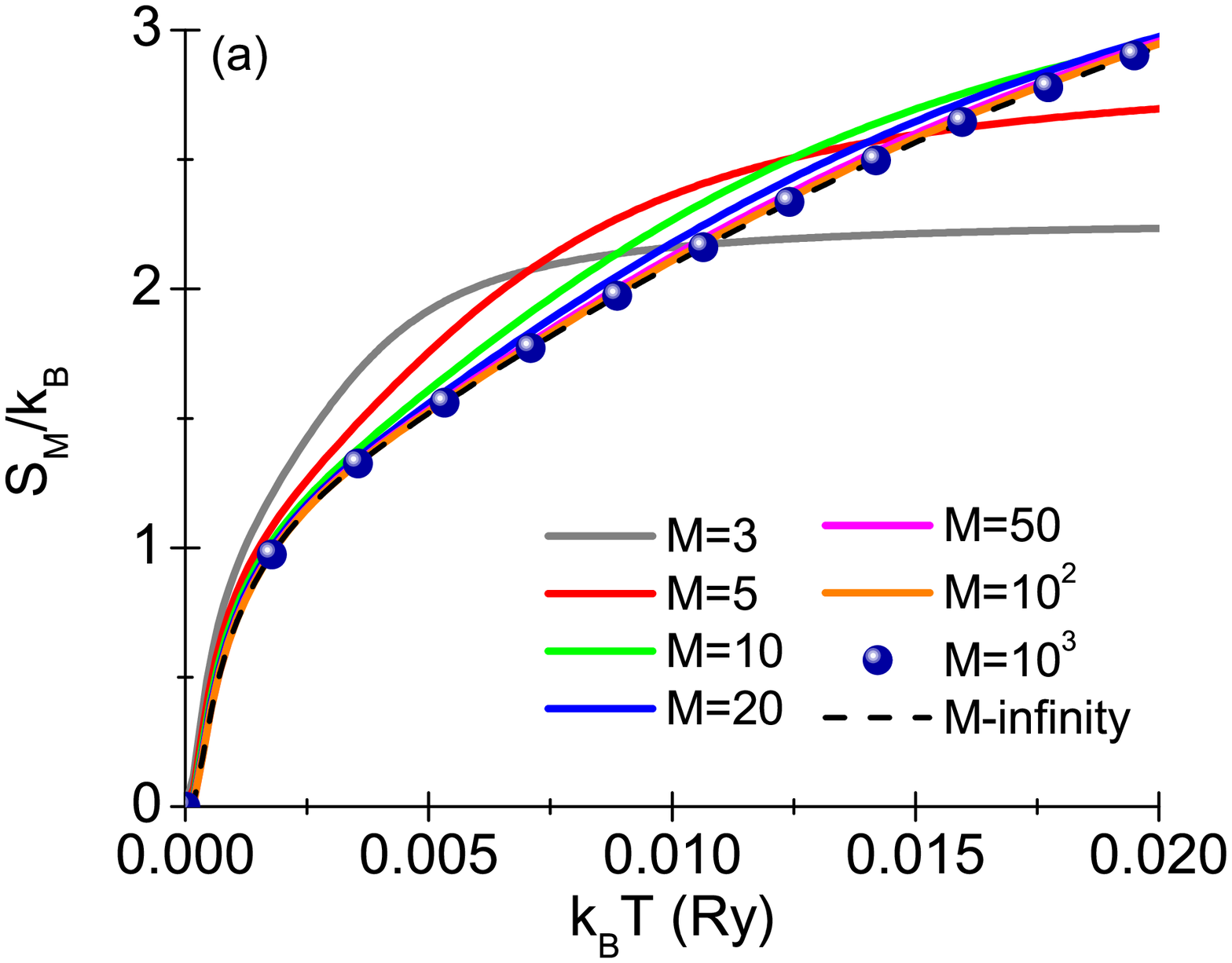}
\includegraphics[width=0.49\columnwidth]{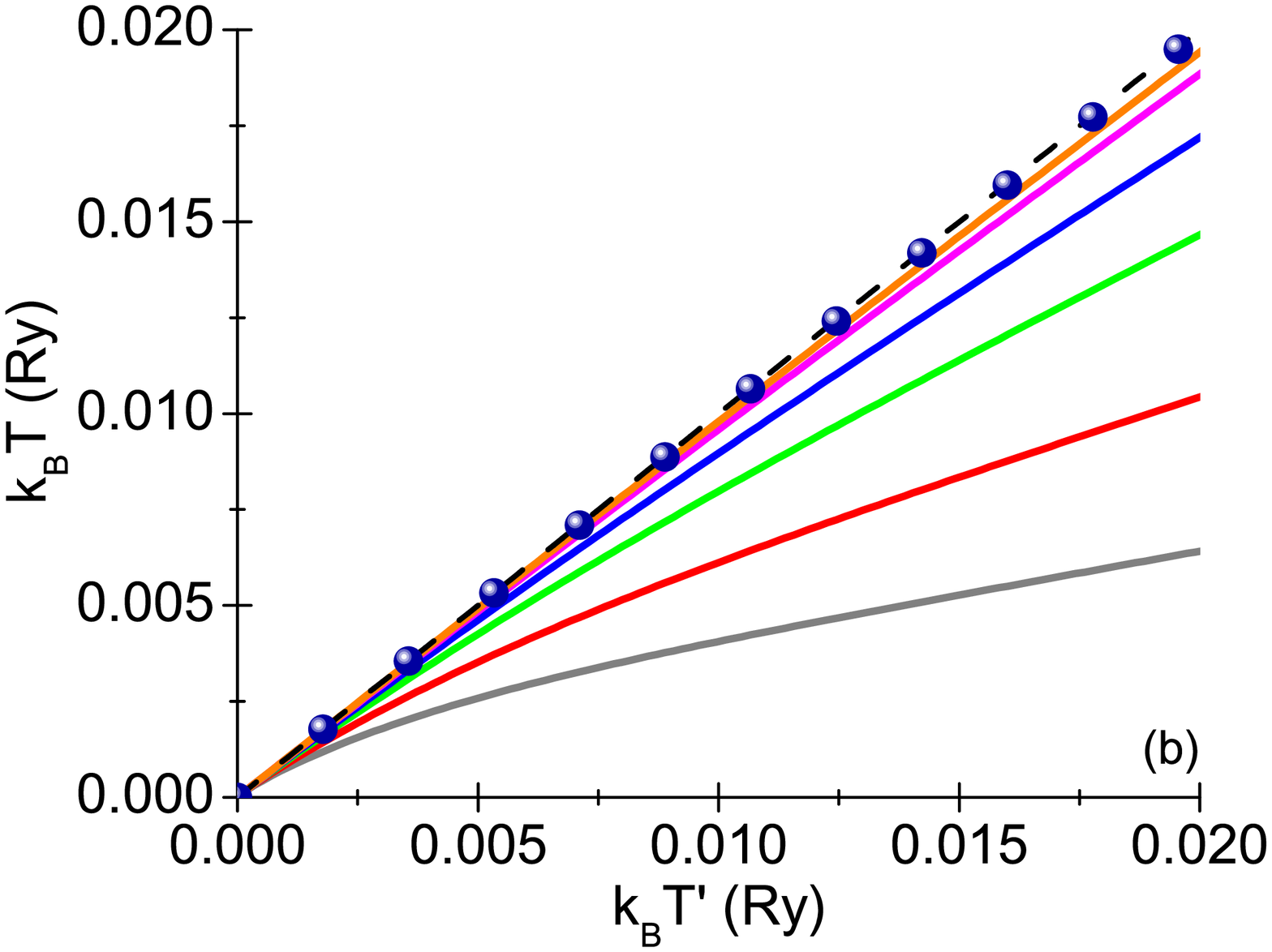}
\caption{(a) The entropy per hydrogen molecule as a function of the temperature for the selected values of $M$. 
         (b) The dependencies between temperatures $T$ and $T'$ obtained using the equation \eq{r03a}.}
\label{f3}
\end{figure}
\begin{figure} 
\includegraphics[width=0.49\columnwidth]{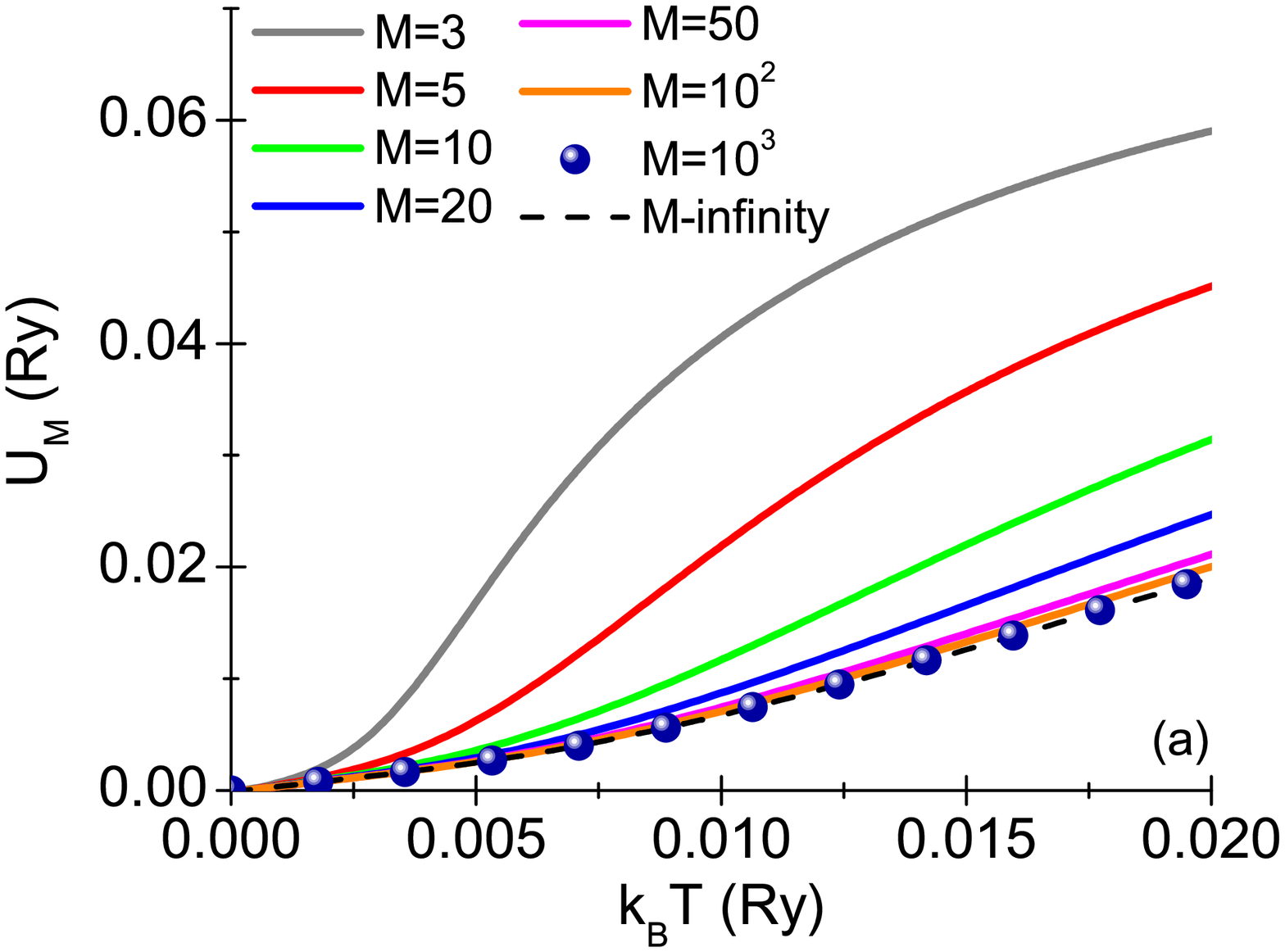}
\includegraphics[width=0.49\columnwidth]{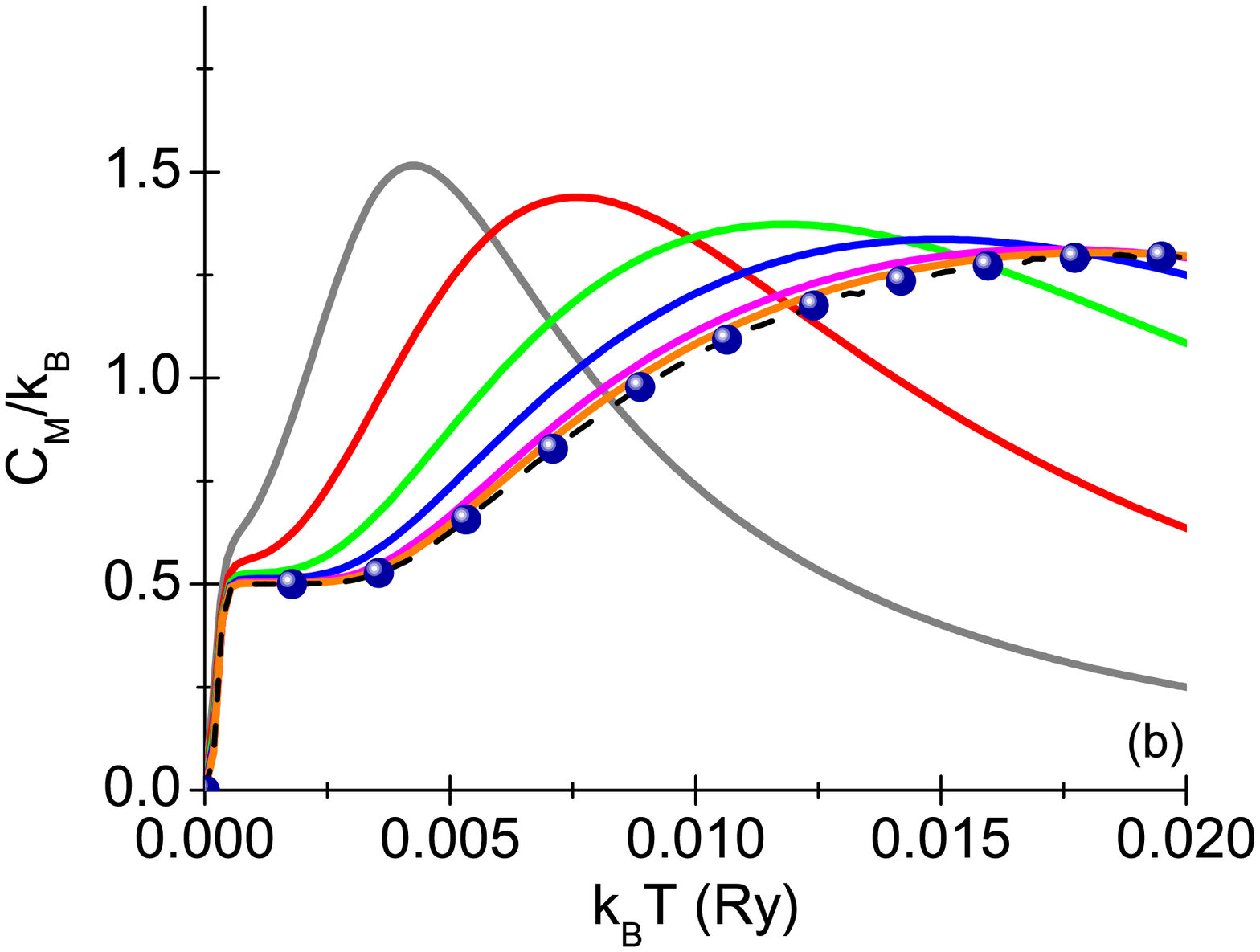}
\caption{(a) The internal energy and (b) the specific heat per hydrogen molecule as a function of the temperature for the selected values of $M$.}
\label{f4}
\end{figure}

In the first step, we took into account the "frozen" system (in the considerations we skip $E_{t}$) and the oscillating energy calculated  in the harmonic approximation. We have adopted the reasonable range of the values of the thermal energy $k_{B}T$ from $0$ to 
$\sim 0.02$~Ry, which corresponds to the temperature at $3000$~K (in each case the value of $T$ was calculated using the equation \eq{r03a}). The above assumptions mean that the hydrogen molecule is in the ground electronic state and because of the value of the dissociation energy ($E_{d}=0.323011$~Ry) has at most $12$ oscillatory states and $24$ rotational states. Of course, not all combinations of the oscillatory and rotational states will be realized physically due to the fact that $E_{o}+E_{r}<E_{d}$. 

\fig{f3} (a) presents the influence of the temperature on the value of the entropy for selected $M$. It can be seen that 
the course of the entropy strongly depends on the number of the particles in the system. According to the presented data, it is further shown that for the case of $M_{c}\sim 10^{3}$ molecules, the differences between the values of $S_{M_{c}}$ and $S_{+\infty}$ are already very slight in the entire temperature range. It means that the nonextensivity of the entropy vanishes quite quickly with the increase in the number of the particles. \fig{f3} (b) shows the plot of the dependence $T\left(T'\right)$. For all the analyzed values of $M$ we have obtained curves monotonically growing, which from the physical point of view is the expected and desired outcome.   

\fig{f4} (a) and (b) presents the curves of the internal energy and the specific heat ($C_{M}=\partial U_{M}/\partial T$). The obtained results show that the values of $U_{M}$ attributable to the single molecule are the smallest in the thermodynamic limit. In the case of the specific heat, the curves $C_{M}\left(T\right)$ have more complicated course than the curves of the internal energy, however, below the value of $M_{c}$ the deviations from the predictions of the classical statistical physics are easily noticeable.

\begin{figure} 
\includegraphics[width=0.49\columnwidth]{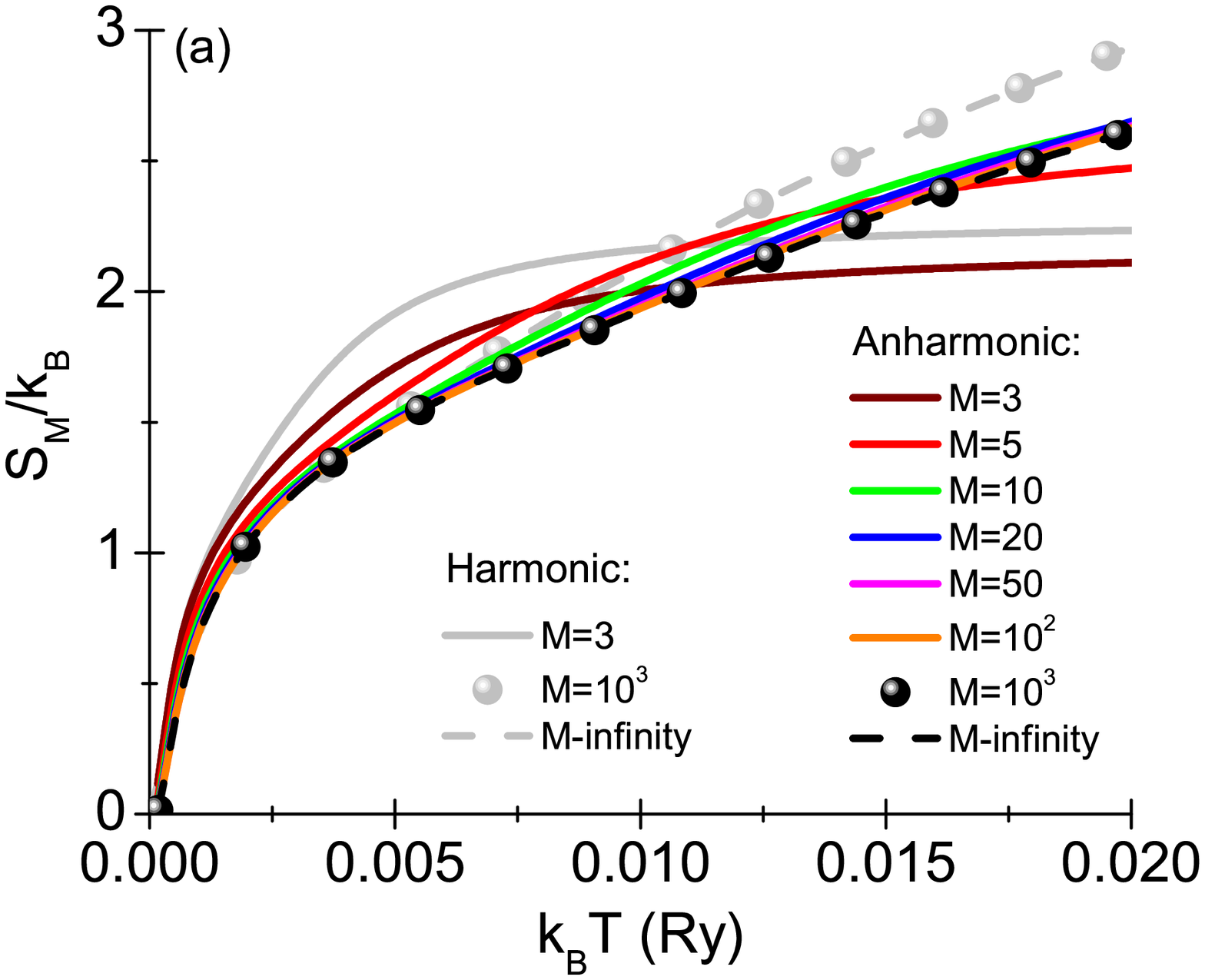}
\includegraphics[width=0.49\columnwidth]{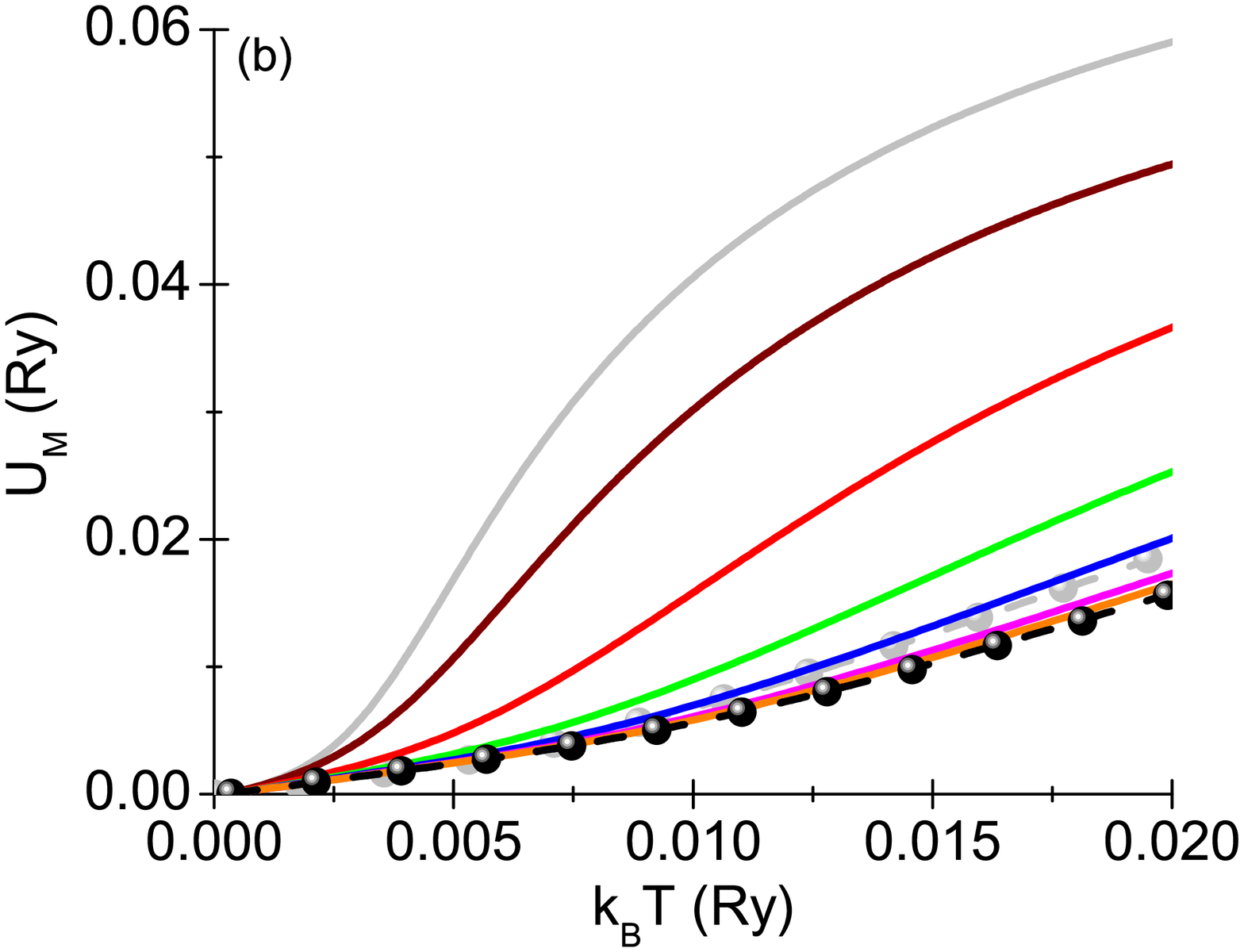}
\includegraphics[width=0.49\columnwidth]{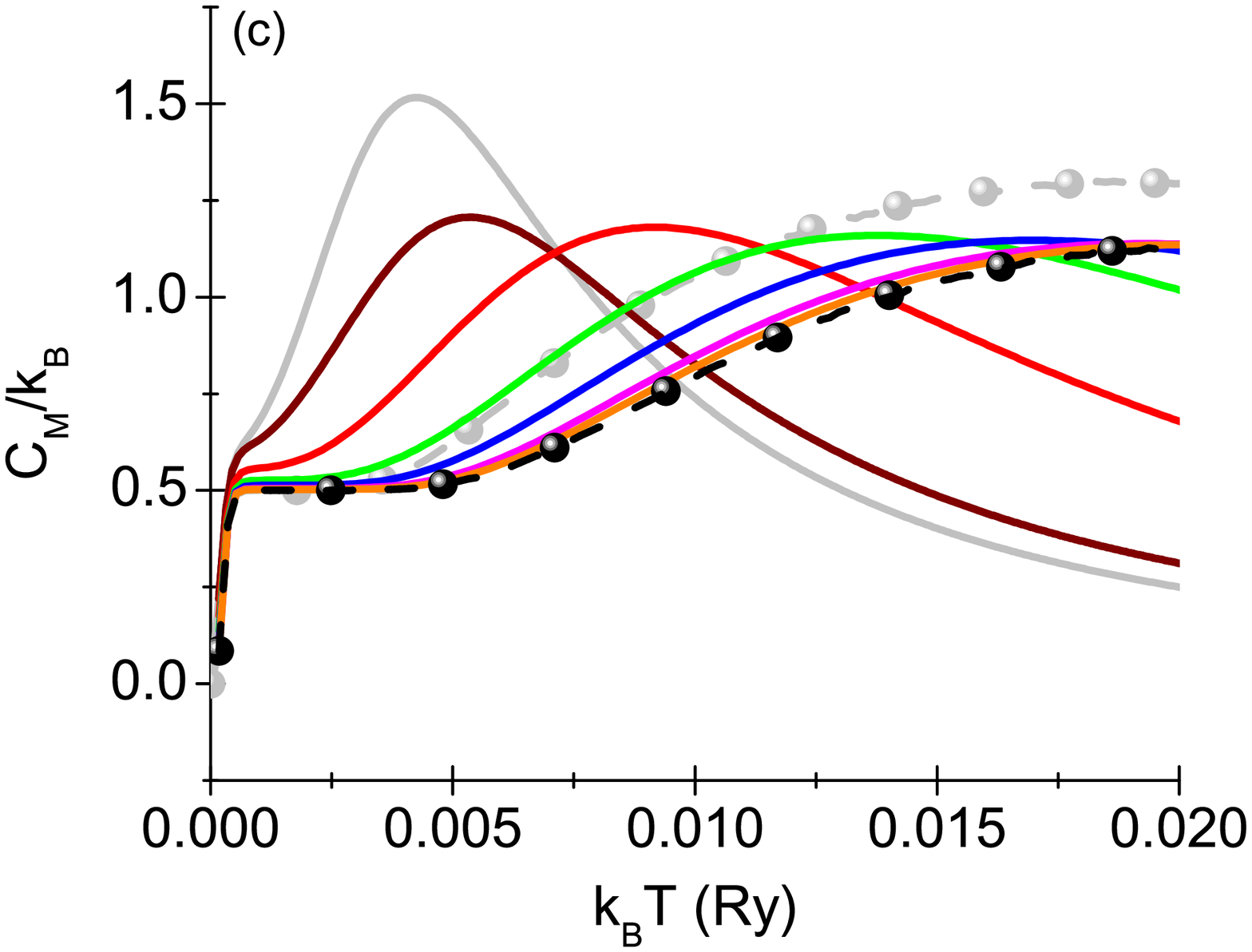}
\includegraphics[width=0.49\columnwidth]{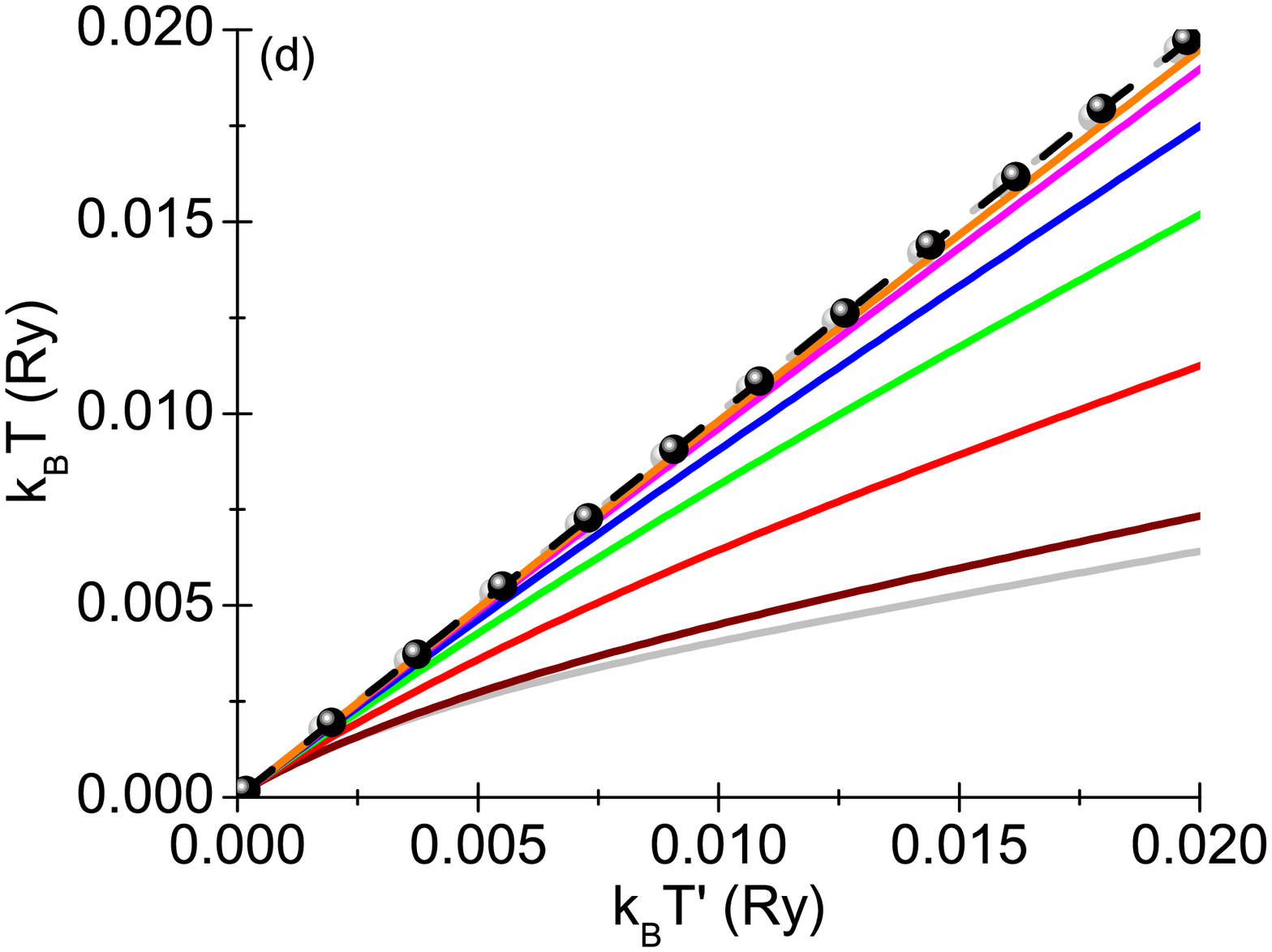}
\caption{The values of the entropy (a), the internal energy (b), the specific heat (c), and the temperature (d) for the selected values of $M$. The main results were obtained taking into account the anharmonicity of the vibrations of the hydrogen molecule. In the background, we have placed the results for the harmonic case.}
\label{f5}
\end{figure}

The results presented above have been obtained in the approximation of the harmonic vibrations (formula \eq{r13b}). 
It is possible to ask the natural question whether the change in the oscillatory energy levels of the hydrogen molecule obtained as part of the anharmonic approach (formula \eq{r15b}) will noticeably affect the non-extensiveness of the system under study. The sample results for the entropy, the internal energy, the specific heat, and the temperature were collected in \fig{f5} (a)-(d). It can be seen that in the anharmonic case, the parameter $M_{c}$ does not change its value, however, as expected, the values of the calculated thermodynamic quantities change. The obtained result is the strong argument suggesting that the value of the parameter $M_{c}$ characterizes the universal feature of the small-scale statistical systems which is non-extensiveness. Especially that the same result as in the anharmonic case is obtained taking into account the change in the rotational levels of the hydrogen molecule induced by the centrifugal force (then instead of the formula \eq{r17b} it is necessary to use formula: 
$E_{r}=B_{0}l\left(l+1\right)-B_{1}l^{2}\left(l+1\right)^{2}+B_{2}l^{3}\left(l+1\right)^{3}$). 

\begin{figure} 
\includegraphics[width=0.49\columnwidth]{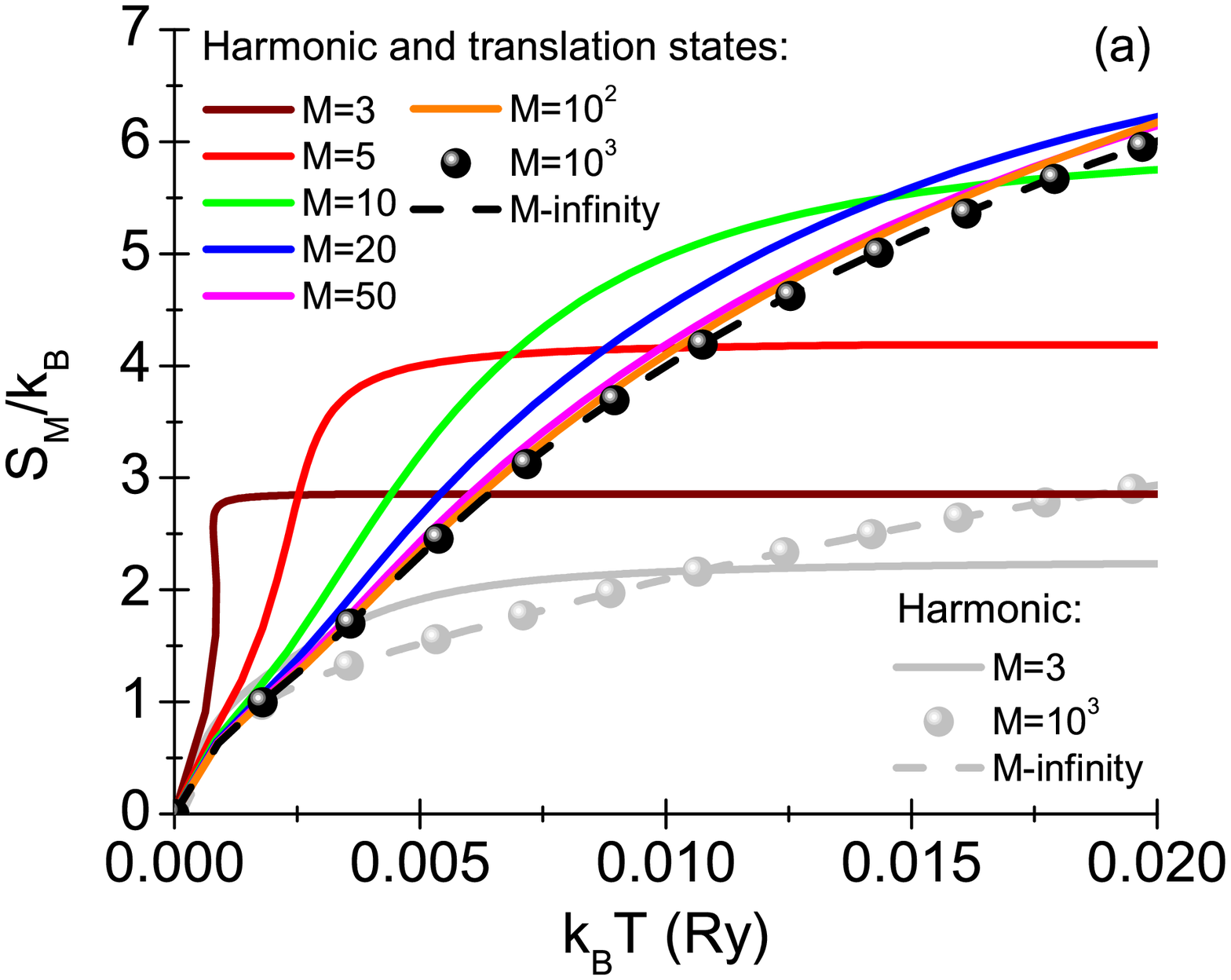}
\includegraphics[width=0.49\columnwidth]{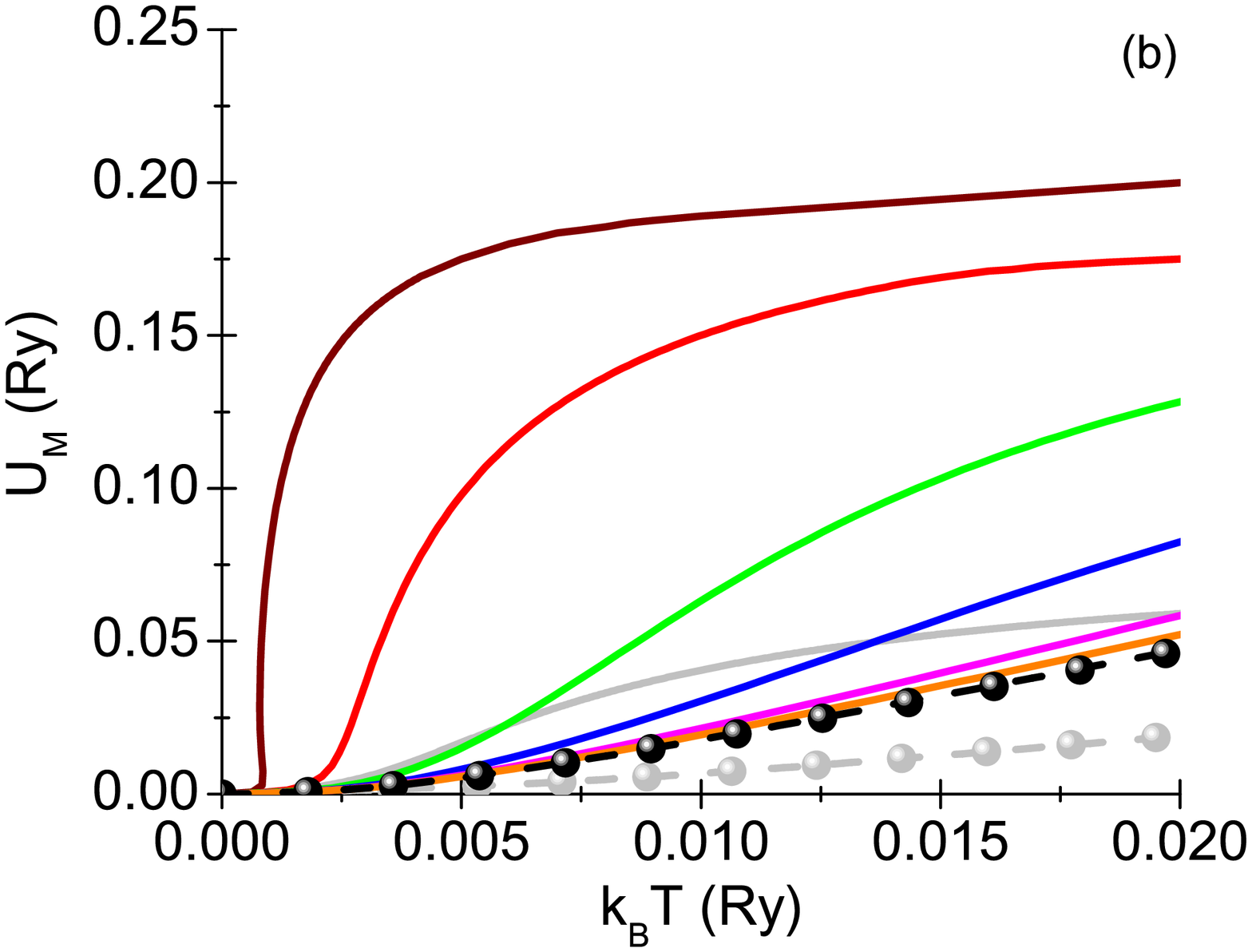}
\includegraphics[width=0.49\columnwidth]{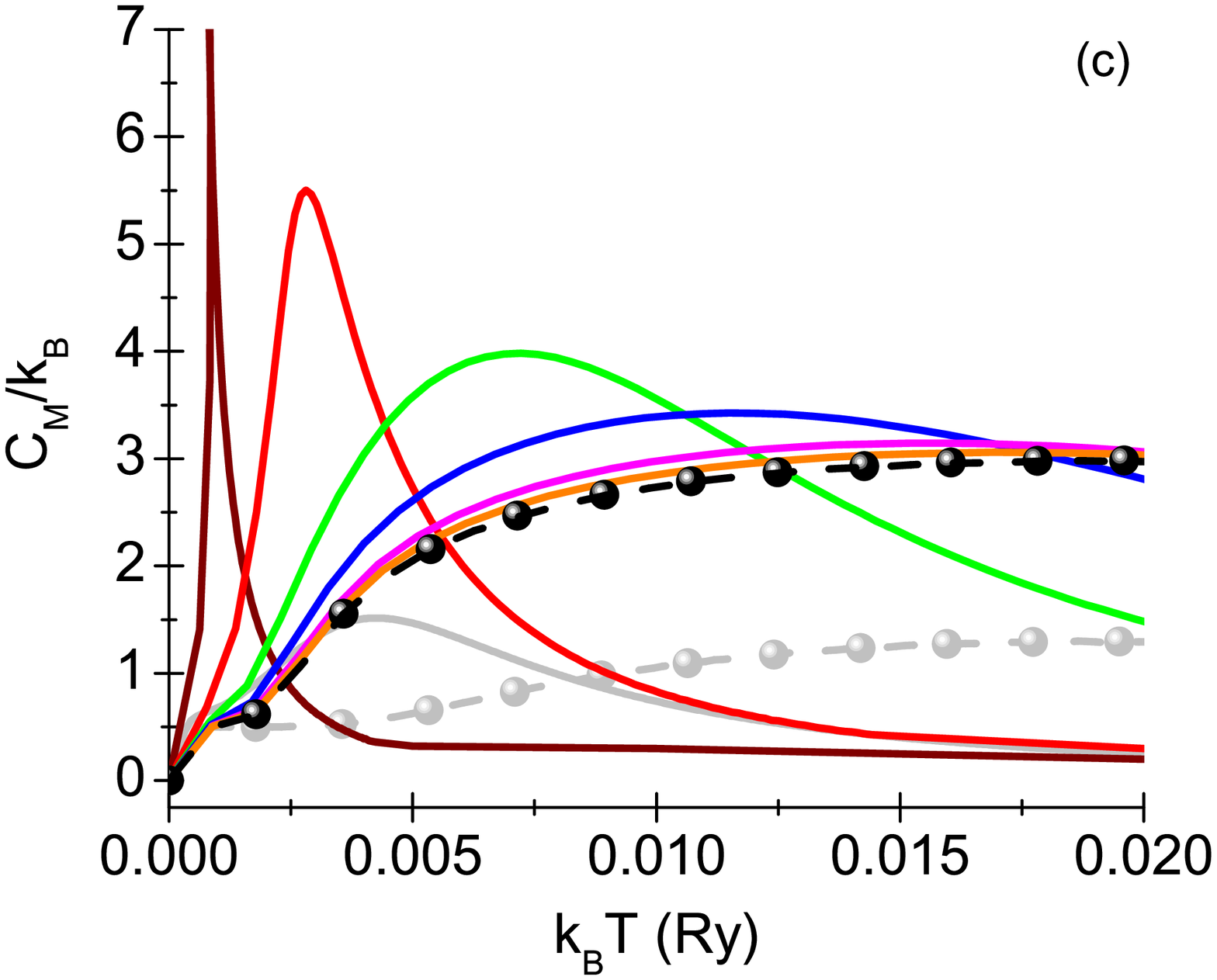}
\includegraphics[width=0.49\columnwidth]{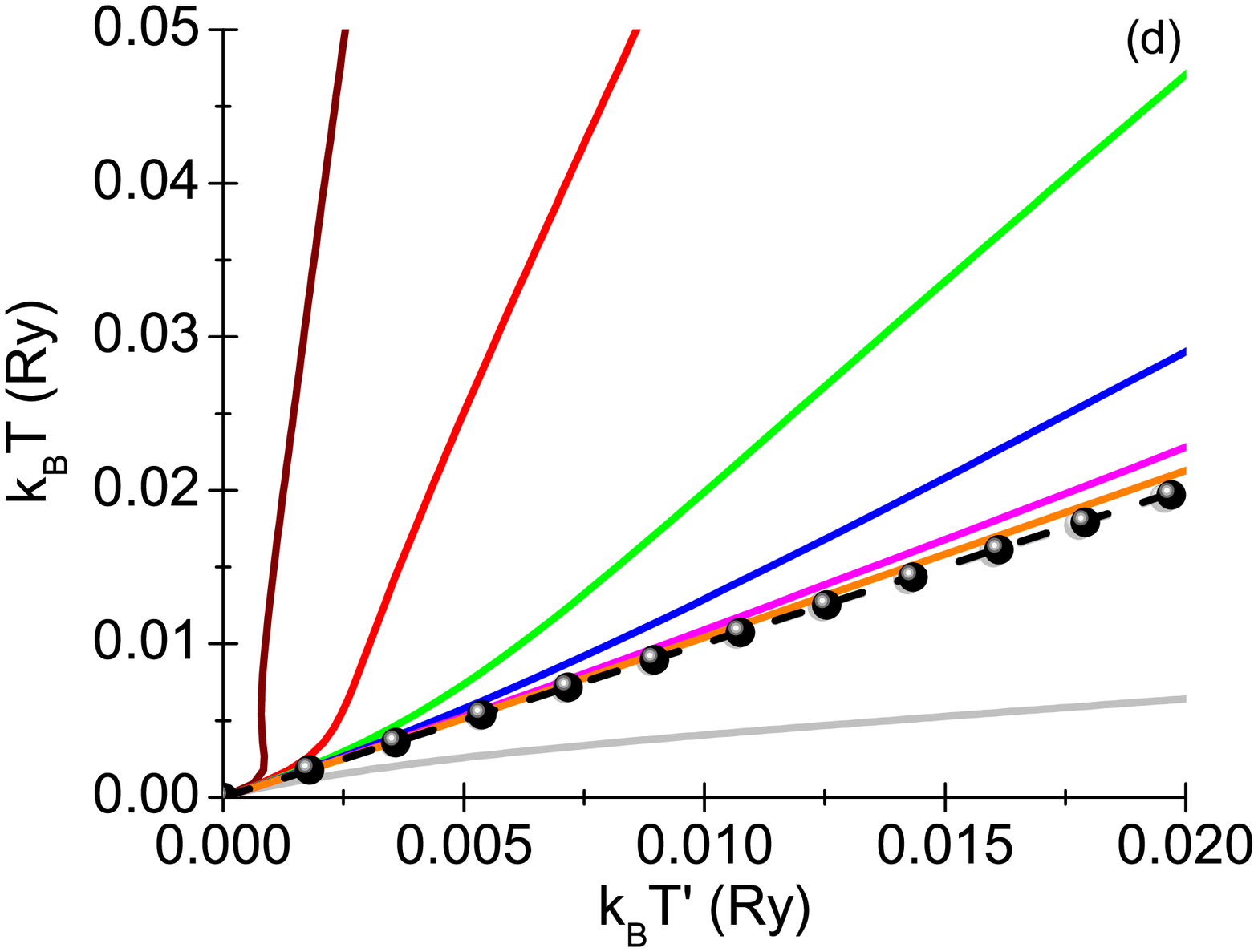}
\caption{The entropy (a), the internal energy (b), the specific heat (c), and the temperature (d) for the selected values of $M$. 
The results from the first plan were obtained for the harmonic case additionally taking into account $\sim 125000$ translational states. In the background, we put the results for the harmonic case while ignoring the translational degrees of freedom.}
\label{f6}
\end{figure}

The change in the value of $M_{c}$ was not observed also when taking into account in the calculations the translational degrees of freedom. The obtained results were collected in \fig{f6}.

\begin{table*}
\caption{\label{t1}  The distance ($R_{0}$), the inverse size of orbital ($\alpha_{0}$), 
                     the ground-state energy ($E_{0}$) or the enthalpy ($H_{0}$) if $F\neq 0$, the vibrational energies 
                     ($\omega^{\rm H}_{0}$, $\omega^{\rm Mo}_{0}$), and the rotational constant ($B_{0}$) 
                     for the equilibrium case and the different values of $F$.}
\begin{ruledtabular}
\begin{tabular}{|c||c|c|c|c|c|c|}
                       &                         &                                   &                 &     &     &     \\
$F$ (Ry/${\rm a_{0}}$) & $R_{0}$ (${\rm a_{0}}$) & $\alpha_{0}$ (${\rm a^{-1}_{0}}$) & $E_{0}$ or $H_{0}$ (Ry)   & $\omega^{\rm H}_{0}$ (${\rm Ry}$) 
& $\omega^{\rm Mo}_{0}$ (${\rm Ry}$) & $B_{0}$ (${\rm Ry}$)\\
\hline
                         &                &                 &                &                &      &         \\
{\bf 0} & {\bf 1.419680} & {\bf 1.199205} & {\bf -2.323011} & {\bf 0.027449} & {\bf 0.038240} & {\bf 0.000540} \\ 
   0.1  &      1.302372  &      1.226840  &      -2.187303  &  0.033599      & -              & 0.000642       \\
   0.2  &      1.219082  &      1.248111  &      -2.061430  &  0.038806      & -              & 0.000733       \\
   0.3  &      1.154596  &      1.265578  &      -1.942867  &  0.043456      & -              & 0.000817       \\ 
   0.4  &      1.102119  &      1.280465  &      -1.830113  &  0.047720      & -              & 0.000897       \\ 
   0.5  &      1.057994  &      1.293465  &      -1.722167  &  0.051699      & -              & 0.000973       \\ 
        &                &                &                 &                &                &                \\         
\end{tabular}
\end{ruledtabular}
\end{table*}
\begin{table*}
\caption{\label{t2}  The equilibrium values of the Hamiltonian parameters $\hat{H}$ 
                     for the different values of $F$.}
\begin{ruledtabular}
\begin{tabular}{|c||c|c|c|c|c|c|}
                       &                         &                  &                &                &                &                 \\
$F$ (Ry/${\rm a_{0}}$) & $\varepsilon_{0}$ (Ry)  & $t_{0}$ (Ry)     & $U_{0}$ (Ry)   & $K_{0}$ (Ry)   & $J_{0}$ (Ry)   & $V_{0}$ (Ry)    \\
\hline
                       &                         &                  &                &                &                &                 \\
        {\bf 0}        &  {\bf -1.749493}        & {\bf -0.737679}  & {\bf 1.661254} & {\bf 0.962045} & {\bf 0.022040} & {\bf -0.011850} \\
         0.1           &  -1.738782              & -0.834142        & 1.711008       & 1.005356       & 0.023190       & -0.012134       \\
         0.2           &  -1.724986              & -0.913237        & 1.749406       & 1.038483       & 0.024031       & -0.012414       \\
         0.3           &  -1.709850              & -0.981633        & 1.781021       & 1.065607       & 0.024694       & -0.012685       \\ 
         0.4           &  -1.694103              & -1.042553        & 1.808033       & 1.088691       & 0.025242       & -0.012946       \\
         0.5           &  -1.678108              & -1.097854        & 1.831676       & 1.108838       & 0.025708       & -0.013198       \\
                       &                         &                  &                &                &                &                 \\         
\end{tabular}
\end{ruledtabular}
\end{table*}

The analogous calculations we have conducted for the case, when each molecule is affected by the external force. Referring to the quantity $E_{F}$ in the equation \eq{r01b} it can be seen that it is connected directly with the pressure inside the chamber in which the tested system is located. The action of the force $F$ in the range of the values from $0$ to $0.5$~Ry/${\rm a_{0}}$ causes the noticeable decrease in the equilibrium distance of the hydrogen molecule. Decreases also the value of the equilibrium ground-state enthalpy ($H_{0}$). For this reason, the selected numerical data is collected in \tab{t1} and in \tab{t2}.     

\begin{figure*} 
\includegraphics[width=1\columnwidth]{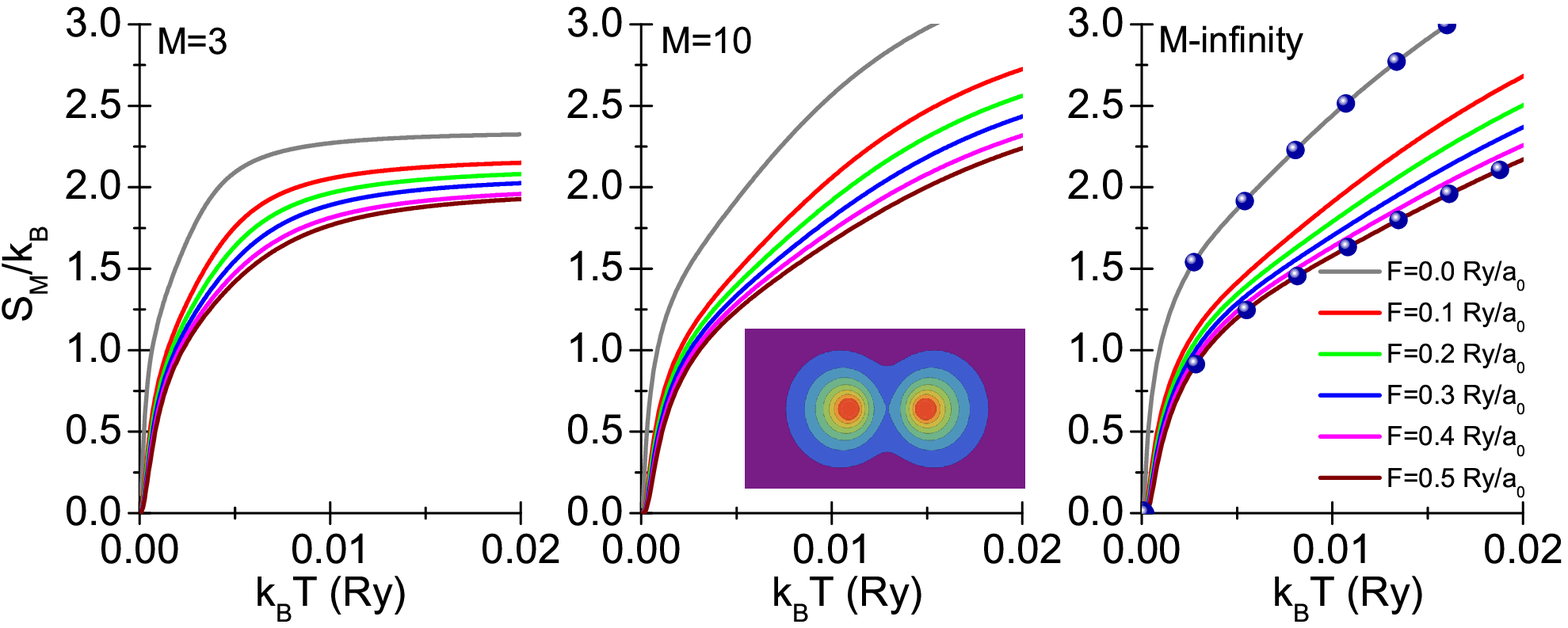}
\caption{(a)-(c) The entropy per hydrogen molecule as a function of the temperature for the selected values of $M$ and $F$. 
On the Fig. 7 (c) points denotes the results obtained for $M=10^{3}$. 
          The insertion presents the distribution of the electron charge density in the equilibrium condition for $F=0.5$~${\rm Ry/a_{0}}$.}
\label{f7}
\includegraphics[width=1\columnwidth]{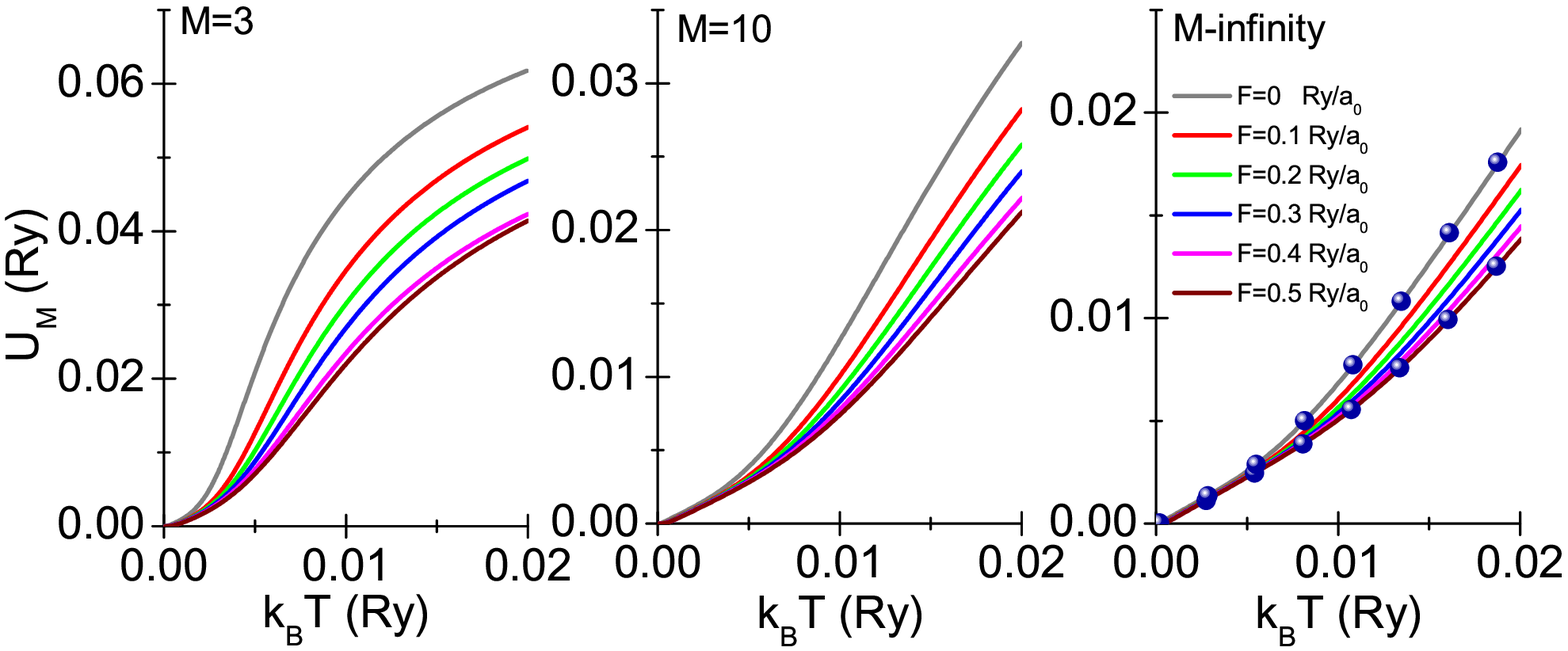}
\caption{(a)-(c) The internal energy per hydrogen molecule as a function of the temperature for the selected values of $M$ and $F$.
         On the Fig. 8 (c) points denotes the results obtained for $M=10^{3}$.}
\label{f8}
\includegraphics[width=1\columnwidth]{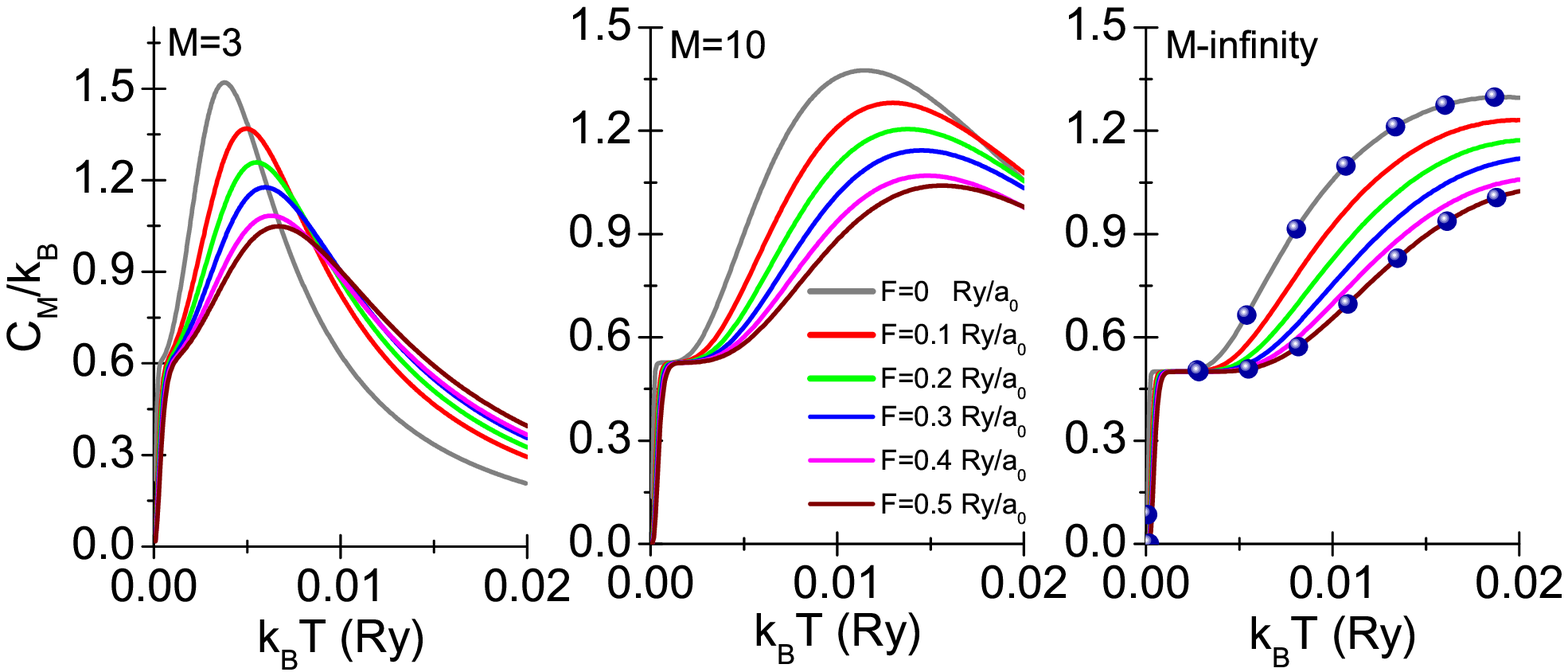}
\caption{(a)-(c) The specific heat per hydrogen molecule as a function of the temperature for the selected values of $M$ i $F$.
        On the Fig. 9 (c) points denotes the results obtained for $M=10^{3}$.}

\label{f9}
\end{figure*}

\fig{f7}-\fig{f9} show the plots of the courses of the entropy, the internal energy, and the specific heat as a function of the temperature 
for the selected values of $F$ and $M$. The obtained results prove that the increasing force $F$ changes visibly the values of the considered thermodynamic functions. However, the adopted value of $M_{c}$ is not subjected to change.

\begin{figure} 
\includegraphics[width=0.49\columnwidth]{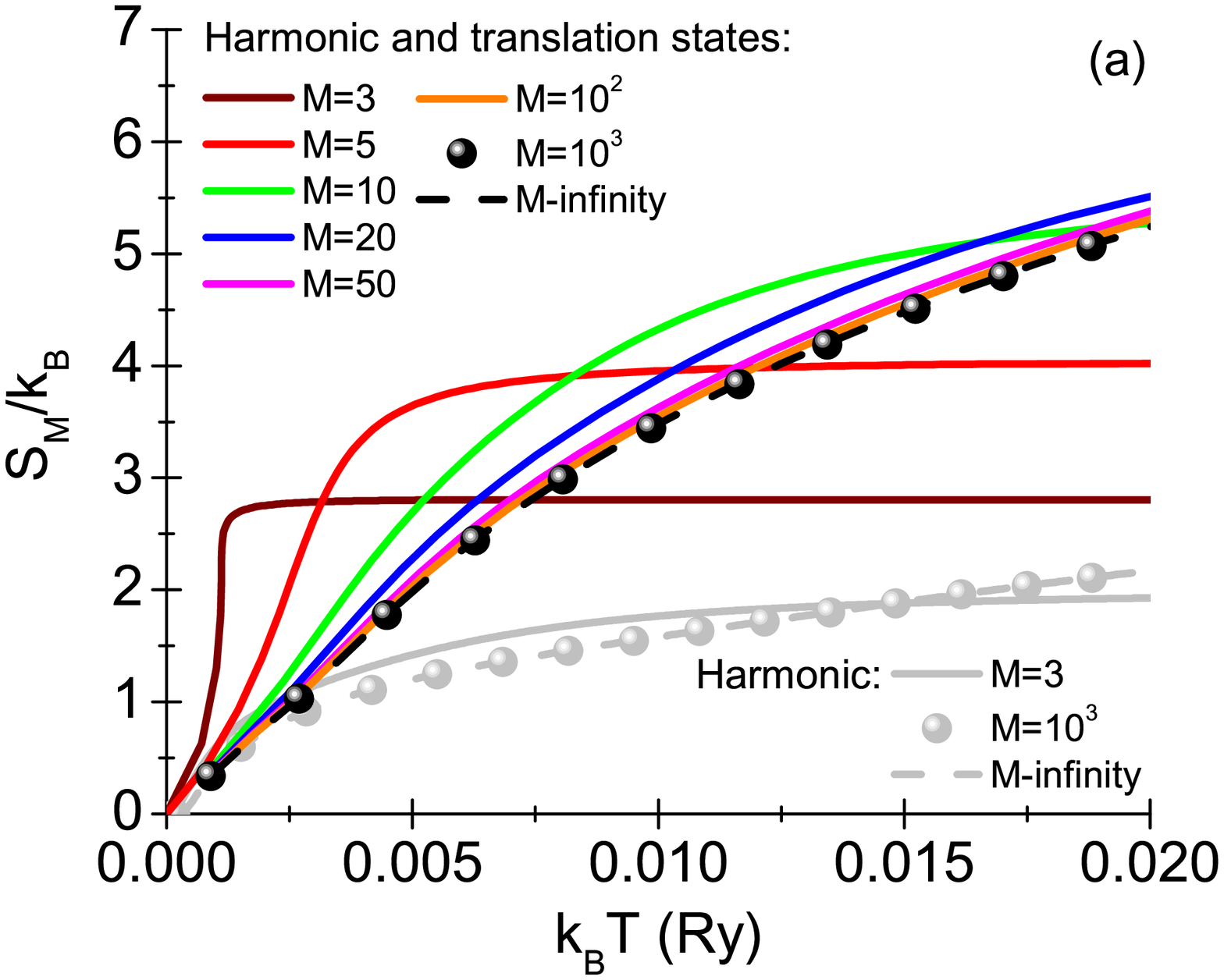}
\includegraphics[width=0.49\columnwidth]{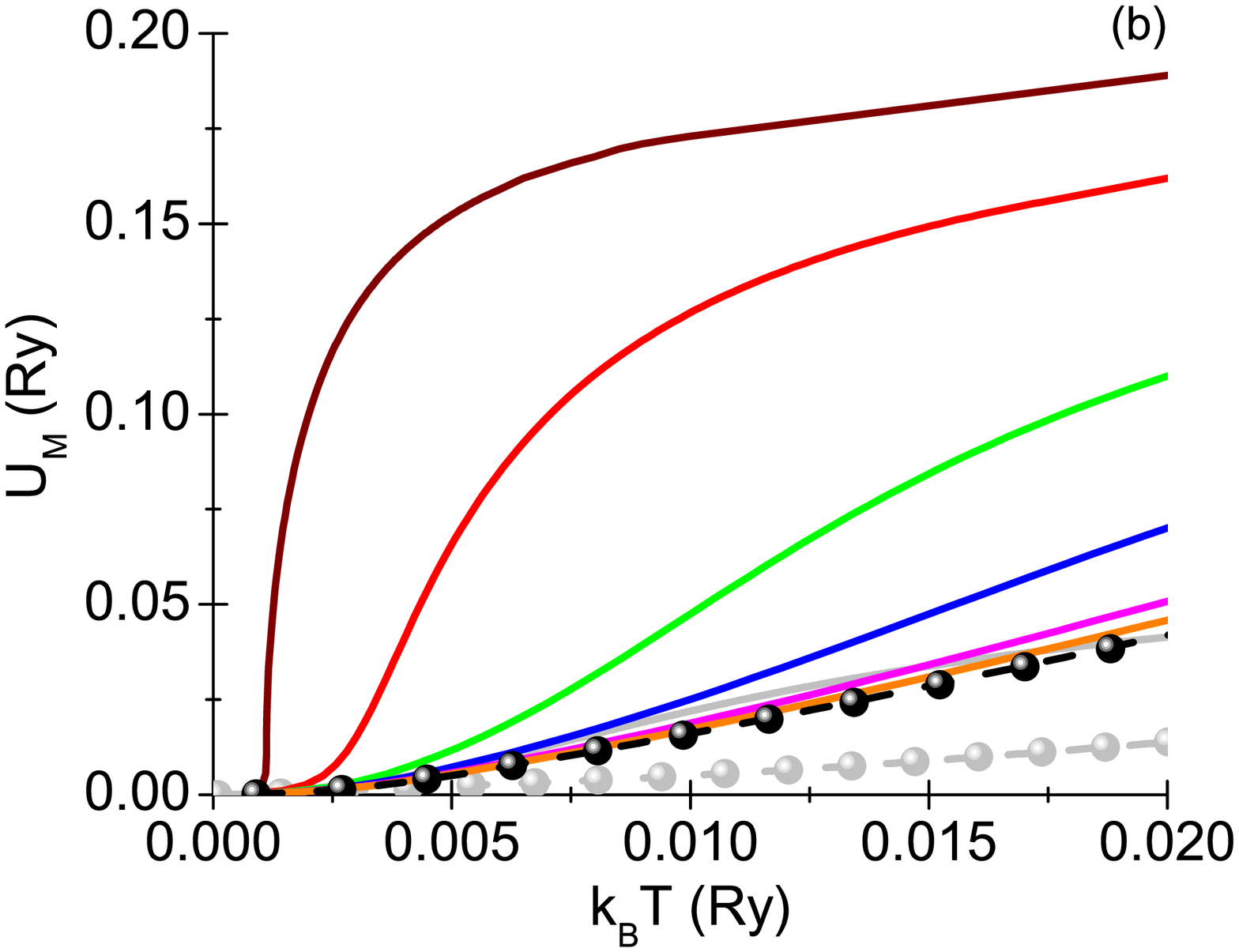}
\includegraphics[width=0.49\columnwidth]{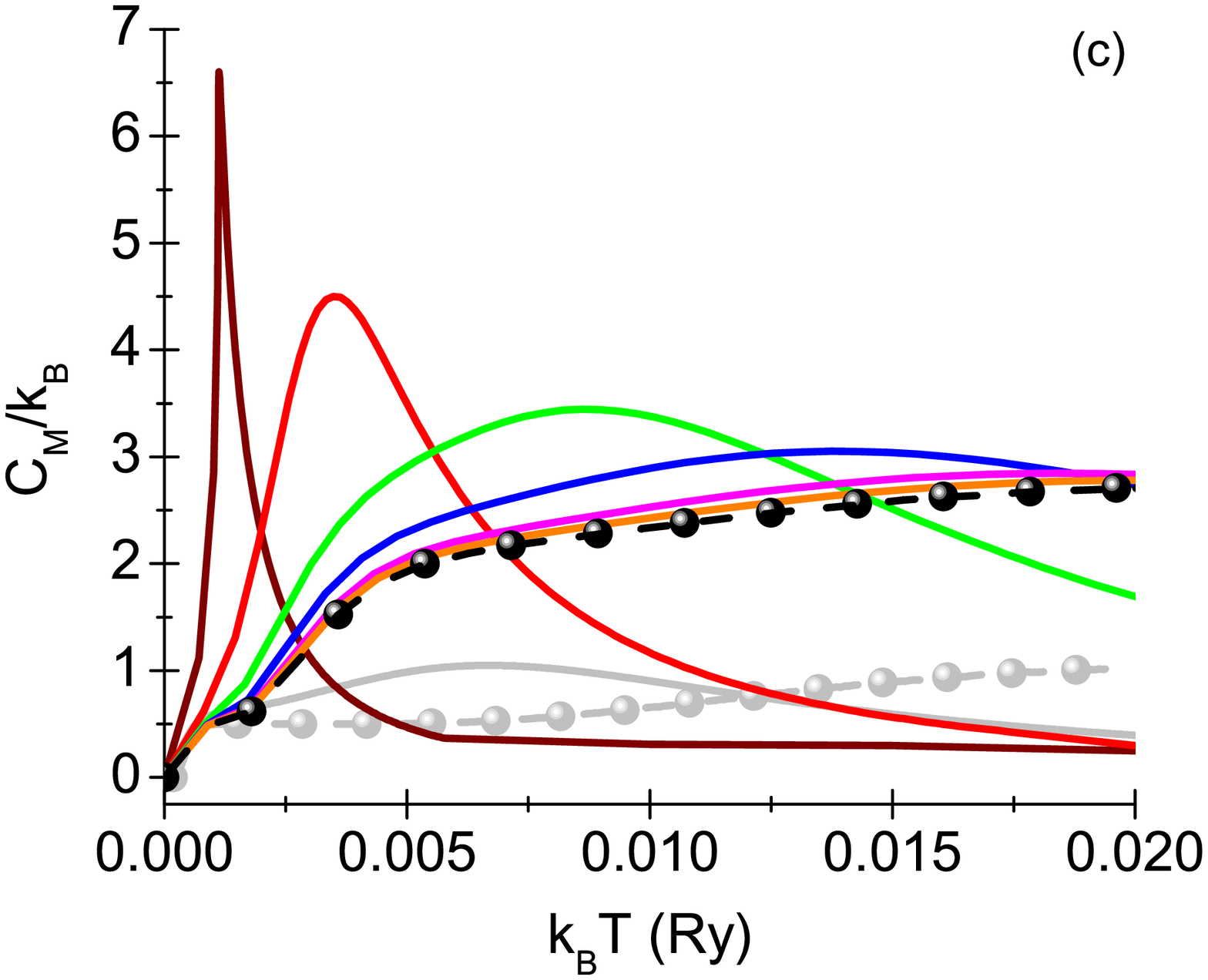}
\includegraphics[width=0.49\columnwidth]{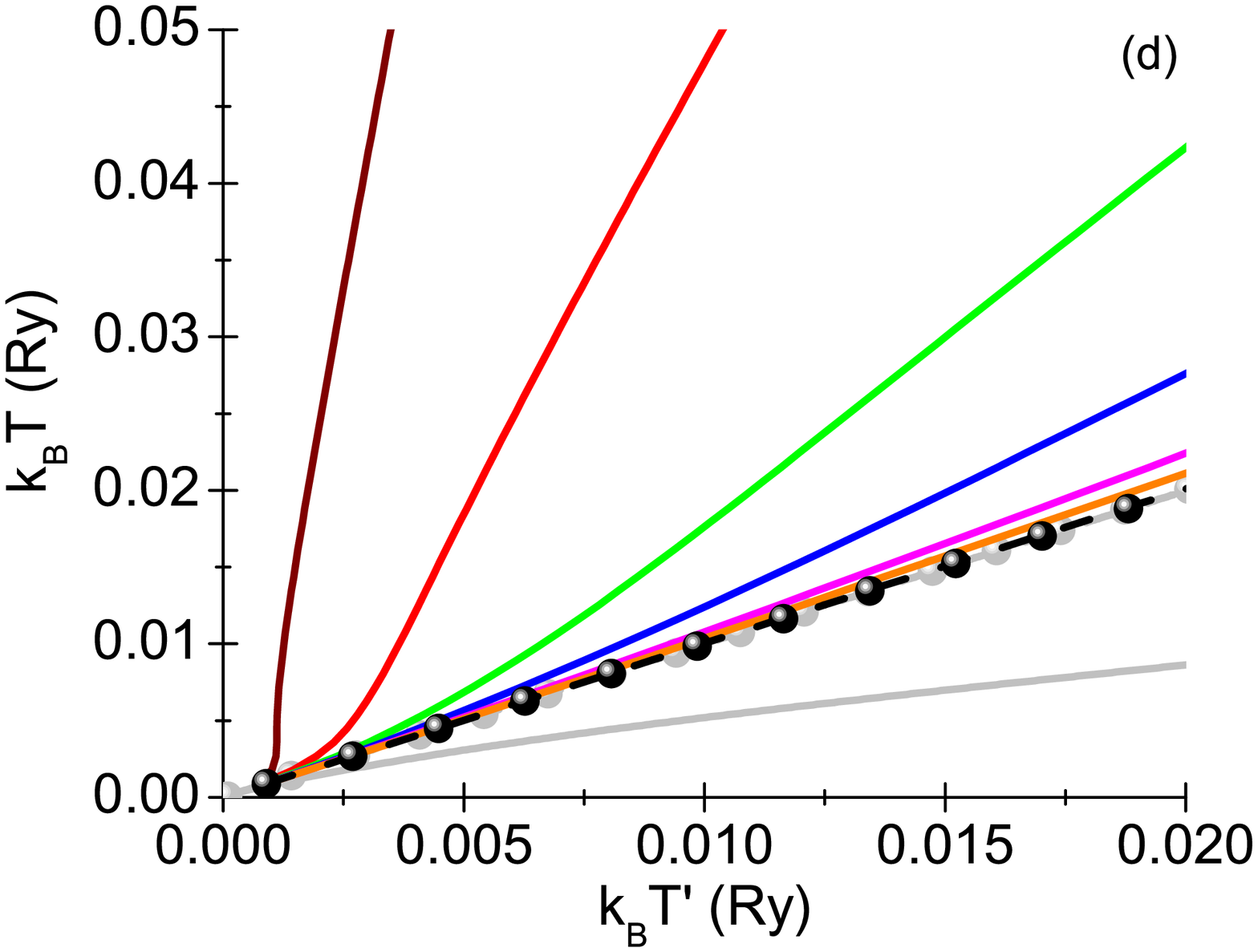}
\caption{The entropy (a), the internal energy (b), the specific heat (c), and the temperature (d) for the selected values of $M$ and 
$F=0.5$~Ry/${\rm a_{0}}$. The main results were obtained for the harmonic case, additionally taking into account $\sim 125000$ translational states. In the background, we put the results for the harmonic case while ignoring the translational degrees of freedom.}
\label{f10}
\end{figure}

\fig{f10} (a)-(d) present the results obtained for the case of the maximum value of the force ($F=0.5$~Ry/${\rm a_{0}}$), while additionally, we took into account the translational degrees of freedom. Analyzing the collected results, one can notice again the lack of change in the value of the parameter $M_{c}$.

In the last part of the chapter, let's note that in addition to the force $F$ affecting the hydrogen molecule, other external disturbance may be related to the external magnetic field $h$. It turns out that in this case neither the value of $M_{c}$ nor the values of the thermodynamic parameters $S_{M}$, $U_{M}$ and $C_{M}$ change. This result can be easily understood considering that for the magnetic fields smaller than $h_{c}=\dfrac{1}{4}\left(K-U-2J+\sqrt{N}\right)=0.586399$~Ry (the field value causing the degeneration of the electronic ground level ($E_{5}=E_{1}$)), only one level of the energy changes in the physically available energy ladder ($E_{1}$).

\section{The discussion of the results}

The analysis conducted in the presented paper has shown that in the systems consisted of the hydrogen molecules, it is possible to observe the significant deviation from the predictions of the classical statistical physics unless the system consists of the number of the molecules less than thousand ($M_{c}\sim 10^{3}$). Physically, it means that the value of $M_{c}$ marks the universal boundary between the non-extensive statistical physics of Tsallis and the theory of Boltzmann-Gibbs-Shannon at least in the considered family of the systems. This is evidenced by the results obtained for the case of the harmonic and anharmonic vibrations of the hydrogen molecule. Additionally, we checked that $M_{c}$ does not change when we take into account the influence of the centrifugal force on the rotational energy levels, or we take into account the translational degrees of freedom. What is important, the value of $M_{c}$ also does not change when the external constant force affects the molecule, or the molecule is located in the external magnetic field. However, except for the magnetic field with reasonable values, shapes of the determined thermodynamic functions are changing. The obtained results are even more convincing because they were obtained as part of the non-parametric model. 

In our opinion, the presented results should be of the particular interest to those involved in the description of the thermodynamic properties of the nanosystems, which by definition are small-sized statistical objects. For future research related to the non-extensive physics one of the most interesting directions of the research is linked with the interactions between the particles. However, this requires the generalization of the nonextensive formalism in the scheme presented in the works \cite{Fetter1971A, Elk1979A}.

\bibliography{Bibliography}
\end{document}